\documentclass[preprint,aps,showpacs,nofootinbib,preprintnumbers,amsmath,amssymb,superscriptaddress]{revtex4-1}
\usepackage{amssymb}
\usepackage{epsfig}
\usepackage{latexsym}
\usepackage{graphicx}
\usepackage{subfigure}
\usepackage{dcolumn}
\usepackage{bm}
\usepackage[usenames ,dvipsnames]{xcolor}
\usepackage[utf8]{inputenc}
\usepackage{slashed}
\usepackage{cleveref}

\begin{document}
\title{Unitarity Bounds on the Massive Spin-2 Particle Explanation of Muon $g-2$ Anomaly}
\author{Da~Huang\footnote{dahuang@bao.ac.cn}}
\affiliation{National Astronomical Observatories, Chinese Academy of Sciences, Beijing, 100012, China}
\affiliation{School of Fundamental Physics and Mathematical Sciences, Hangzhou Institute for Advanced Study, UCAS, Hangzhou 310024, China}
\affiliation{International Centre for Theoretical Physics Asia-Pacific, Beijing/Hangzhou, China}
\author{Chao-Qiang~Geng\footnote{cqgeng@ucas.ac.cn}}
\affiliation{School of Fundamental Physics and Mathematical Sciences, Hangzhou Institute for Advanced Study, UCAS, Hangzhou 310024, China}
\affiliation{International Centre for Theoretical Physics Asia-Pacific, Beijing/Hangzhou, China}
\author{Jiajun~Wu\footnote{wujiajun@itp.ac.cn}}
\affiliation{School of Fundamental Physics and Mathematical Sciences, Hangzhou Institute for Advanced Study, UCAS, Hangzhou 310024, China}
\affiliation{International Centre for Theoretical Physics Asia-Pacific, Beijing/Hangzhou, China}

\date{\today}
\begin{abstract}
Motivated by the long-standing discrepancy between the Standard Model prediction and the experimental measurement of the muon magnetic dipole moment, we have recently proposed to interpret this muon $g-2$ anomaly in terms of the loop effect induced by a new massive spin-2 field $G$. In the present paper, we investigate the unitarity bounds on this scenario. We calculate the $s$-wave projected amplitudes for two-body elastic scatterings of charged leptons and photons mediated by $G$ at high energies for all possible initial and final helicity states. By imposing the condition of the perturbative unitarity, we obtain the analytic constraints on the charged-lepton-$G$ and photon-$G$ couplings.  We then apply our results to constrain the parameter space relevant to the explanation of the
muon $g-2$ anomaly.    
\end{abstract}

\maketitle

\section{Introduction}\label{s1}
One of the greatest puzzles in the Standard Model (SM) is the discrepancy between the SM theoretical prediction and experimental data on the muon magnetic dipole moment $(g-2)_\mu$~\cite{pdg}, which is regarded as a hint towards  new physics beyond the SM (BSM). Currently, by combining the data from Brookhaven~\cite{Muong-2:2006rrc} and Fermilab~\cite{Muong-2:2021ojo}, the muon $g-2$ anomaly is given by
\begin{eqnarray}
	\Delta a_\mu = a_\mu^{\rm exp} -  a_\mu^{\rm SM} = (25.1 \pm 5.9) \times 10^{-10}\,,
\end{eqnarray}
where the latest SM calculation leads to $a_\mu^{\rm SM} = (116591810\pm 43)\times 10^{-11}$~\cite{Aoyama:2020ynm,Keshavarzi:2018mgv,Chao:2021tvp}. Already, many BSM models have been proposed in order to resolve this muon $g-2$ discrepancy (for a recent review see {\it e.g.}~\cite{Athron:2021iuf} and references therein).

In Ref.~\cite{Huang:2022}, we have demonstrated  that the muon $g-2$ anomaly can be explained by the one-loop effects induced by a new spin-2 particle $G$, which can be identified as the first Kaluza-Klein (KK) graviton in the generalized Randall-Sundrum (RS) model~\cite{Davoudiasl:1999tf,Pomarol:1999ad,Chang:1999nh,Davoudiasl:2000wi,Batell:2005wa,Batell:2006dp,Fok:2012zk,Lee:2013bua,Han:2015cty,Geng:2016xin,Falkowski:2016glr,Dillon:2016fgw,Dillon:2016tqp,Kraml:2017atm,Geng:2018hpq,Goyal:2019vsw}. By calculating all relevant one-loop Feynman diagrams, we have obtained the analytic expression of the leading-order $G$-induced contribution to the muon $g-2$, which is shown to maintain both the gauge invariance of the quantum electrodynamics and the correct divergence structure of loop integrals. Note that we have only imposed the theoretical constraint from the perturbativity on the parameter space of the spin-2 particle model in Ref.~\cite{Huang:2022}. However, another criterion to determine if the perturbative calculation remains under control is the perturbative unitarity bound, which has a long history to restrict parameters in a given model~\cite{Gell-Mann:1969cuq,Weinberg:1971fb}. Perhaps the most famous application was to limit the mass of the SM Higgs boson in Ref.~\cite{Lee:1977yc,Lee:1977eg,Durand:1989zs}. An incomplete list of the use of unitarity bounds to restrict the BSM models includes Refs.~\cite{Glashow:1976nt, Huffel:1980sk, Maalampi:1991fb, Kanemura:1993hm, Akeroyd:2000wc, Das:2015mwa, Kanemura:2015ska, Goodsell:2018tti, Appelquist:1987cf, Chaichian:1987zt, Falkowski:2016glr,Banta:2021dek}. 

In the present work, we would like to derive the unitarity bounds for the spin-2 particle model in Ref.~\cite{Huang:2022}. Note that the interpretation of the muon $g-2$ anomaly involves $G$ couplings to charged leptons $\ell$ and photons $\gamma$. Thus, in order to constrain these two kinds of interactions, one needs to consider the 2-to-2 elastic scatterings of charged leptons and of photons via the mediation of $G$, with all possible initial and final helicity states. By yielding the $s$-wave projected amplitudes for all scattering processes, we can impose the unitarity bounds on the model parameters. As a result, it will be shown that the obtained unitarity bounds have significant impacts on the parameter space to explain the muon $g-2$.  

The paper is organized as follows. In Sec.~\ref{model}, we briefly summarize the main results in Ref.~\cite{Huang:2022}, including the Lagrangian of the spin-2 particle model and the leading-order expression of the lepton $g-2$ contribution induced by $G$. Sec.~\ref{SecUniBound} is devoted to the calculation of amplitudes for elastic scattering processes, $\ell^- \ell^+ \to \ell^- \ell^+$ and $\gamma\gamma\to \gamma\gamma$, with all possible initial and final helicity states. From these amplitudes, we derive the unitarity bounds on the charged-lepton- and photon-$G$ couplings. We then make advantage of the obtained unitarity bounds to constrain the muon $g-2$ preferred parameter space in Sec.~\ref{SecResult}. Finally, we conclude in Sec.~\ref{Conclusion}.  

\section{A Spin-2 Particle Model and its Explanation to Lepton $g-2$}\label{model}
The Lagrangian for the spin-2 particle $G$ explanation to the lepton $g-2$ is given by~\cite{Han:1998sg}
\begin{eqnarray}\label{LagG}
    {\cal L}_G = -\frac{1}{\Lambda} G_{\mu\nu} \Big[c_\gamma T_\gamma^{\mu\nu} + \sum_{\ell=e,\mu,\tau} c_\ell T_\ell^{\mu\nu}\Big]\,,
\end{eqnarray}
where the stress-energy tensors, $T_\ell^{\mu\nu}$ and $T_\gamma^{\mu\nu}$, of charged leptons  and  photons  are defined by
\begin{eqnarray}\label{DefEMT}
    T_\ell^{\mu\nu} &=& \frac{i}{4} \bar{\ell} (\gamma^\mu \partial^\nu + \gamma^\nu \partial^\mu) \ell -\frac{i}{4} (\partial^\mu \bar{\ell}\gamma^\nu + \partial^\nu \bar{\ell}\gamma^\mu) \ell \nonumber\\
    && -i\eta^{\mu\nu} [\bar{\ell} \gamma^\rho \partial_\rho \ell + im_\ell \bar{\ell}\ell -\frac{1}{2}\partial^\rho (\bar{\ell}\gamma_\rho \ell)]\,, \nonumber\\
    T_\gamma^{\mu\nu} &=& \frac{1}{4} \eta^{\mu\nu} F^{\rho\sigma}F_{\rho\sigma} - F^{\mu\rho}F^{\nu}_\rho \nonumber\\
    && -\frac{1}{\xi} \left[\eta^{\mu\nu}\left(\partial^\rho \partial^\sigma A_\sigma A_\rho + \frac{1}{2}(\partial^\rho A_\rho)^2\right)-\left(\partial^\mu \partial^\rho A_\rho A^\nu + \partial^\nu \partial^\rho A_\rho A^\mu\right)\right]\,,
\end{eqnarray}
respectively, with $\xi$ the gauge parameter for the photon field. This Lagrangian can be viewed as a part of the low-energy effective action in the generalized Randall-Sundrum model where the massive spin-2 particle is the first KK excitation of the conventional graviton~\cite{Randall:1999ee,Davoudiasl:1999tf,Pomarol:1999ad,Chang:1999nh,Davoudiasl:2000wi,Falkowski:2016glr}. 

In the light of the effective interactions between $G$ and charged leptons/photons in Eq.~(\ref{LagG}), we have drawn and calculated in Ref.~\cite{Huang:2022} the one-loop Feynman diagrams contributing to the charged lepton $g-2$, with the dominant contribution given by
\begin{eqnarray}\label{G2total}
	\Delta a_{\ell}^G = \left(\frac{m_\ell^2}{\Lambda}\right)^2 \left(\frac{\Lambda}{m_G}\right)^4 \left(\frac{c_\ell^2}{48\pi^2} - \frac{c_\ell c_\gamma}{24\pi^2}\right)\,.
\end{eqnarray}
When deriving this leading-order result, we have applied the loop regularization~\cite{Wu:2002xa,Wu:2003dd} method to regularize the quartic divergences of loop integrals, which has been shown to maintain the gauge invariance of quantum electrodynamics and the correct divergence structure simultaneously. 

Moreover, by requiring the validity of the perturbative expansion, {\it i.e.}, the one-loop contributions to the lepton $g-2$ should dominate over the the two-loop ones, we obtain novel constraints from the perturbativity on our non-renormalizable spin-2 particle interactions as follows
\begin{eqnarray}
	|c_\ell| < 4\pi \left(\frac{m_G}{\Lambda}\right)^2 \,,\quad \quad	|c_\gamma| < 4\pi \left(\frac{m_G}{\Lambda}\right)^2 \,,
\end{eqnarray}  
which are obviously natural but non-trivial generalizations of perturbativity constraints on renormalizable operators~\cite{Nebot:2007bc}.

\section{Unitarity Bounds}\label{SecUniBound}
In this section, we  
apply the tree-level unitarity bounds as our criterion to determine whether our perturbative calculations  remain under control, which can give extra constraints to our spin-2 particle model. Concretely, the unitarity of the S-matrix imposes the following bound to the $s$-wave projected amplitude $a_0(\sqrt{s})$~\cite{Lee:1977eg,Goodsell:2018tti,Banta:2021dek}
\begin{eqnarray}\label{UniBound}
	{\rm Re}(a_0)(\sqrt{s}) \leq \frac{1}{2}\,,
\end{eqnarray}
where $a_0(\sqrt{s})$ is defined as
\begin{eqnarray}\label{DefA0}
	a_0(\sqrt{s}) = \sqrt{\frac{4|{\bf p}_i| |{\bf p}_f|}{2^{\delta_i+\delta_f}s}} \frac{1}{32\pi} \int^1_{-1} d (\cos\theta) {\cal M}(i\to f) = \sqrt{\frac{4|{\bf p}_i| |{\bf p}_f|}{2^{\delta_i+\delta_f}s}} \frac{1}{16\pi} \int^0_{-s} \frac{d t}{s} {\cal M}(i\to f) \,,
\end{eqnarray}
in which the indices $\delta_{i,f}=1$ if the two particles in the initial or final states are identical to each other, otherwise $\delta_{i,f}=0$.  
In our model, we have two kinds of effective vertices: the photon and lepton couplings to the spin-2 particle $G$, so we need to compute the amplitudes for $\ell^- \ell^+ \to \ell^- \ell^+$ and $\gamma\gamma \to \gamma\gamma$ of various helicity assignments to determine their respective bounds. 

\subsection{$\ell^- \ell^+ \to \ell^- \ell^+$}
\begin{figure}[!htb]
	\centering
	\hspace{-5mm}
	\includegraphics[width=0.49\linewidth]{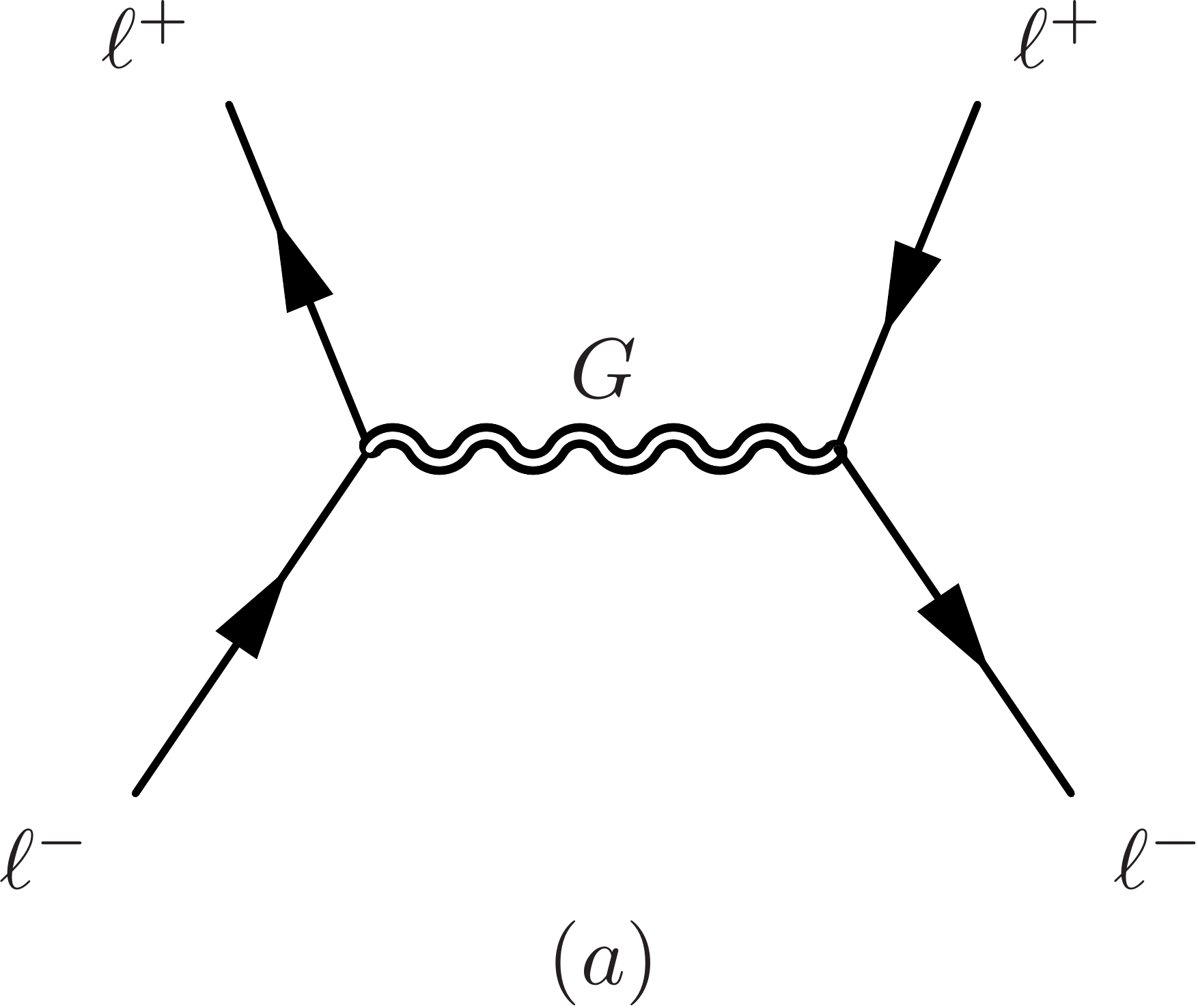}
	\includegraphics[width=0.49\linewidth]{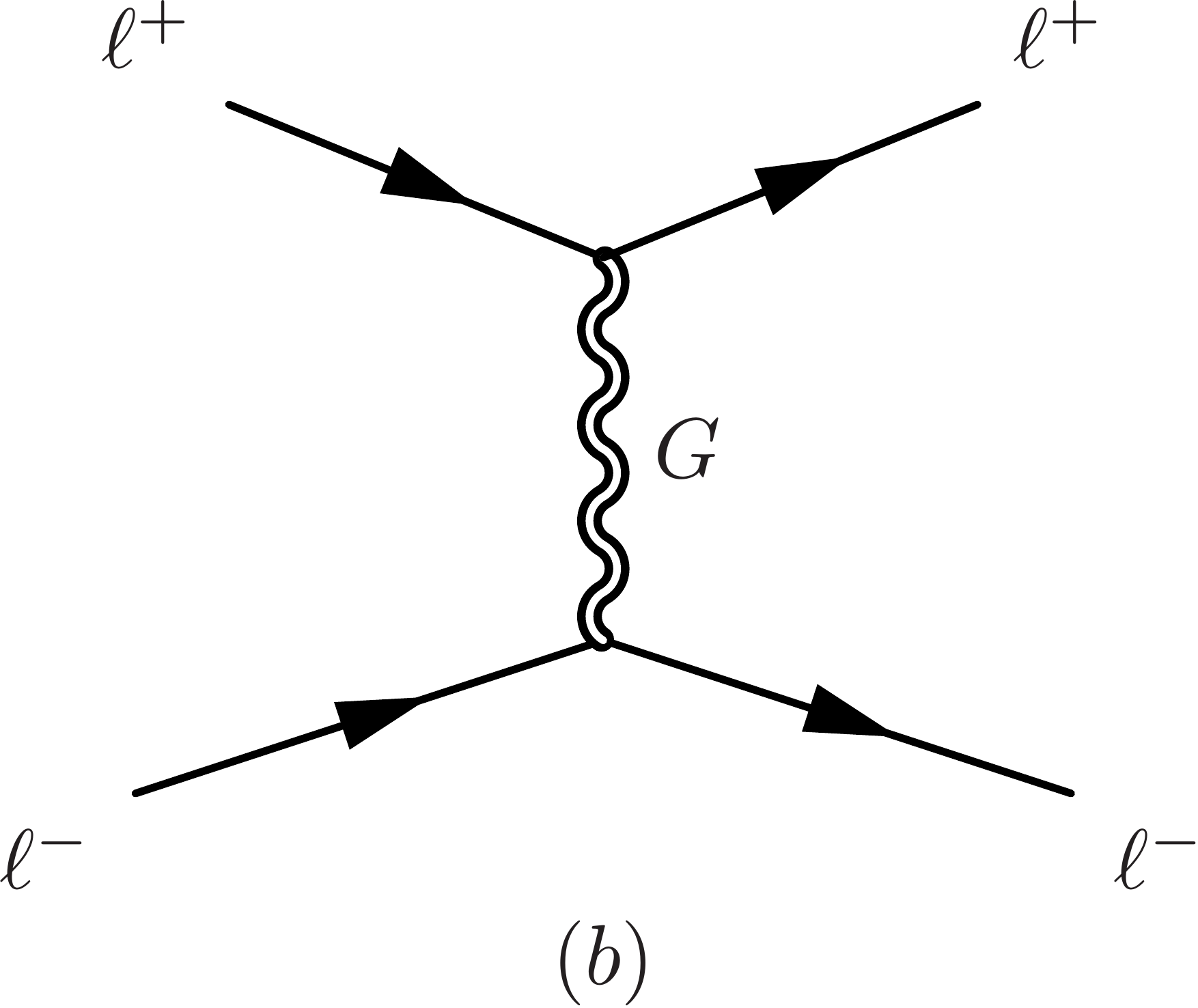}
	\caption{Feynman diagrams which give rise to the scattering process $\ell^- \ell^+ \to \ell^- \ell^+$.}\label{Figll}
\end{figure}

We begin by considering the perturbative unitarity bounds for the lepton-$G$ couplings $c_{\ell}$ with $\ell = e$, $\mu$ and $\tau$, which can be derived from the amplitudes of $\ell^- \ell^+ \to \ell^- \ell^+ $ of different $\ell^\pm$ helicity configurations. Given the Lagrangian in Eq.~(\ref{LagG}), there are two Feynman diagrams contributing to this process, which are shown in Fig.~\ref{Figll} with left and right panels corresponding to the $s$- and $t$-channels, respectively. 
Note that the lepton-$G$ interaction in Eq.~(\ref{LagG}) is $P$-invariant, {\it i.e.}, if the two scatterings of different helicity configurations are related by the parity ${P}$ transformation, then they would be equal in amplitude. For example, we have the following relations
\begin{eqnarray}
	{\cal M} (\ell_R^- \ell^+_L \to \ell^-_R \ell^+_L) = {\cal M} (\ell^-_L \ell^+_R \to \ell^-_L \ell^+_R) \,, \quad
	{\cal M} (\ell_R^- \ell^+_R \to \ell^-_R \ell^+_R) = {\cal M} (\ell^-_L \ell^+_L \to \ell^-_L \ell^+_L) \,.
\end{eqnarray}
Also, we will work in the high-energy limit in which the lepton masses can be ignored compared with the external particle momenta. 
In what follows of this subsection, we shall present the detailed calculations of independent non-zero $\ell^- \ell^+ \to \ell^-\ell^+$ helicity amplitudes. By applying the unitarity bound on these amplitudes, we can obtain constraints on $c_\ell$.  

\subsubsection{$\ell^-_R \ell^+_L \to \ell^-_R \ell^+_L$}
Let us begin by computing the amplitude of $\ell^-_R \ell^+_L \to \ell^-_R \ell^+_L$. In the massless limit, the momenta of external particles in the center-of-mass (com) frame are given by:
\begin{eqnarray}
	&{\rm Incoming}:&\quad	\ell^-_R: \quad p_1 = (E,0,0, E)\,, \quad \quad \ell^+_L: \quad p_2 =(E,0,0,-E)\, \\
	& {\rm Outgoing}:&\quad	\ell^-_R: \quad k_1 = (E,E\sin\theta, 0, E\cos\theta)\,, \quad \quad \ell^+_L:\quad k_2 = (E, -E\sin\theta,0,-E\cos\theta)\,, \nonumber
\end{eqnarray}  
while the polarization vectors are denoted as
\begin{eqnarray}
	&{\rm Incoming}&:	u_R(p_1) = \sqrt{2E} \left(\begin{array}{c}
		0\\
		0\\
		1\\
		0
	\end{array}\right)\,, \quad 
	v_L(p_2) = \sqrt{2E} \left(\begin{array}{c}
		0\\
		0\\
		0\\
		-1\\
	\end{array}\right)\,,\nonumber\\
	&{\rm Outgoing}&: u_R(k_1) = \sqrt{2E} \left(\begin{array}{c}
		0\\
		0\\
		\cos\frac{\theta}{2}\\
		\sin\frac{\theta}{2}
	\end{array}\right)\,, \quad 
	v_L(k_2) = \sqrt{2E} \left(\begin{array}{c}
		0\\
		0\\
		\sin\frac{\theta}{2}\\
		-\cos\frac{\theta}{2}\\
	\end{array}\right)\,.
\end{eqnarray} 

The total amplitude is composed of the $s$- and $t$-channel ones. The $s$-channel amplitude is given by
\begin{eqnarray}
	i{\cal M}_s(\ell_R^- \ell^+_L \to \ell^-_R \ell^+_L) &&= -\frac{ic_\ell^2}{32\Lambda^2} \bar{v}_L (p_2) [\gamma^\mu (p_{1}^\nu - p_2^\nu) + \gamma^\nu (p_1^\mu - p_2^\mu) - 2\eta^{\mu\nu} (\slashed{p}_1 - \slashed{p}_2 - 2m_\ell) ] u_R(p_1)	\nonumber\\
	&& \bar{u}_R(k_1)  [\gamma^\rho (k_{1}^\sigma - k_2^\sigma) + \gamma^\sigma (k_1^\rho - k_2^\rho) - 2\eta^{\rho\sigma} (\slashed{k}_1 - \slashed{k}_2 - 2m_\ell) ] v_L(k_2) \frac{B_{\mu\nu, \rho\sigma}}{Q^2-m_G^2}\,.\nonumber\\
\end{eqnarray}
By using the external particles' equations of motion $\slashed{p}_1 u_R(p_1) = m_\ell u_R(p_1) \simeq 0$ and $\slashed{p}_2  v_L(p_2) = - m_\ell v_L(p_2) \simeq 0$ where the symbol $\simeq$ represents the massless limit, the above $s$-channel amplitude can be simplified to
\begin{eqnarray}\label{sllR}
	i{\cal M}_s(\ell_R^- \ell^+_L \to \ell^-_R \ell^+_L) &=& -\frac{ic_\ell^2}{8\Lambda^2} \big\{ (p_1-p_2)(k_1-k_2) [\bar{u}_R(k_1) \gamma^\mu v_L(k_2)] [\bar{v}_L(p_2) \gamma_\mu u_R(p_1)] \nonumber\\
	&& +  [\bar{u}_R(k_1) (\slashed{p}_1 - \slashed{p}_2) v_L(k_2)][\bar{v}_L(p_2)(\slashed{k}_1 -\slashed{k}_2) u_R(p_1)] \big\} \frac{1}{Q^2-m_G^2} \,.
\end{eqnarray} 
Now we turn to compute the fermion current in this $s$-channel $\ell^-_R \ell^+_L \to \ell^-_R \ell^+_L$ process. Note that 
\begin{eqnarray}
	\gamma^0 \gamma^\mu = \left(\begin{array}{cc}
		0 & 1 \\
		1 & 0
	\end{array}\right) \left(\begin{array}{cc}
		0 & {\sigma}^\mu \\
		\bar{\sigma}^\mu & 0
	\end{array}\right) = \left(\begin{array}{cc}
		\bar{\sigma}^\mu & 0 \\
		0 & \sigma^\mu
	\end{array}\right)\,.
\end{eqnarray}
Thus, for a vector-like current there are only two kinds of nonzero chirality configurations for  $(\ell^-, \ell^+)$ pairs: $(L,R)$ and $(R,L)$, while other two configurations like $(L,L)$ and $(R,R)$ are zero. Hence, for a $\ell^-_R \ell^+_L$ pair, the corresponding current given by
\begin{eqnarray}
	\bar{v}_L(p_2) \gamma^\mu u_R(p_1)  &=& (2E) (0,-1) \sigma^\mu \left(\begin{array}{c}
		1 \\
		0
	\end{array}\right) = (2E)(0,-1,-i,0)\,, \nonumber\\
	\bar{v}_L(k_2) \gamma^\mu u_R(k_1) &=& (2E) (\sin \frac{\theta}{2}, -\cos\frac{\theta}{2}) \sigma^\mu \left(\begin{array}{c}
		\cos\frac{\theta}{2} \\
		\sin\frac{\theta}{2}
	\end{array}\right) = (2E) (0, -\cos\theta, -i, \sin\theta)\,,
\end{eqnarray}
so that
\begin{eqnarray}
	[\bar{u}_R(k_1) \gamma^\mu v_L(k_2)] [\bar{v}_L(p_2) \gamma_\mu u_R(p_1)] =-(2E)^2(1+\cos\theta) = 2u\,,
\end{eqnarray}
\begin{eqnarray}
	\bar{v}_L(p_2)  (\slashed{k}_1 - \slashed{k}_2) u_R(p_1) = (2E)^2 \sin\theta = s \sin\theta \,, \nonumber\\
	\bar{u}_R(k_1) (\slashed{p}_1 - \slashed{p}_2) v_L(k_2) = -(2E)^2 \sin\theta = - s \sin\theta\,.
\end{eqnarray}
Consequently, the s-channel amplitude can be given by
\begin{eqnarray}
	i{\cal M}_s = -\frac{4i c_\ell^2}{32 \Lambda^2}\frac{\left(2u(u-t)-s^2\sin^2\theta\right) }{s-m_G^2}= -\frac{ic_\ell^2}{4 \Lambda^2}\frac{u(u-3t)}{s-m_G^2} = -\frac{ic_\ell^2}{4 \Lambda^2}\frac{(s+t)(s+4t)}{s-m_G^2}\,,
\end{eqnarray}
where in the last equality we have used the following relations
\begin{eqnarray}
	s(1+\cos\theta) = - 2u\,,\quad s(1-\cos\theta) = -2t, \quad 4ut = s^2 \sin^2\theta\,.
\end{eqnarray}

Now we come to compute $t$-channel contribution, with the amplitude given by
\begin{eqnarray}
	i{\cal M}_t (\ell^-_R \ell^+_L \to \ell^-_R \ell^+_L) &&= -\frac{i c_\ell^2}{32\Lambda^2} \bar{v}_L(p_2) [\gamma^\mu (-p_{2}^\nu -k_2^\nu)+\gamma^\nu (-p_2^\mu -k_2^\mu)-2\eta^{\mu\nu}(-\slashed{p}_2 - \slashed{k}_2-2m_\ell)]v_L(k_2) \nonumber\\
	&& \bar{u}_R(k_1)[\gamma^\rho (p_1^\sigma+k_1^\sigma)+\gamma^\sigma(p_1^\rho+k_1^\rho)-2\eta^{\rho\sigma}(\slashed{p}_1+\slashed{k}_1-2m_\ell)]u_R(p_1) \frac{B_{\mu\nu, \rho\sigma}}{q^2-m_G^2}\,.
\end{eqnarray}
By using the lepton on-shell conditions, the $t$-channel amplitude can be reduced into
\begin{eqnarray}
	i{\cal M}_t (\ell^-_R \ell^+_L \to \ell^-_R \ell^+_L) &=& \frac{4ic_\ell^2}{32\Lambda^2}\frac{1}{q^2-m_G^2} \big\{ (k_1+p_1)(k_2+p_2) [\bar{u}_R(k_1) \gamma^\mu u_R(p_1)] [\bar{v}_L(p_2) \gamma_\mu v_L(k_2)] \nonumber\\
	&& + [\bar{u}_R(k_1) (\slashed{k}_2+\slashed{p}_2)u_R(p_1) ][\bar{v}_L(p_2)(\slashed{k}_1 + \slashed{p}_1)v_L(k_2)] 
	\big\}  \,.
\end{eqnarray} 
In the center of mass frame, the t-channel amplitude can be further simplified to
\begin{eqnarray}
	i{\cal M}_t (\ell^-_R \ell^+_L \to \ell^-_R \ell^+_L)  =\frac{ic_\ell^2}{4\Lambda^2} \frac{u(u-3s)}{t-m_G^2} = \frac{ic_\ell^2}{4\Lambda^2} \frac{(t+s)(t+4s)}{t-m_G^2}\,.
\end{eqnarray}
Therefore, the total amplitude is given by summing over the $s$- and $t$-channel amplitudes:
\begin{eqnarray}
	i{\cal M}  (\ell^-_R \ell^+_L \to \ell^-_R \ell^+_L) &=& -\frac{ic_\ell^2}{4\Lambda^2} \left[\frac{u(u-3t)}{s-m_G^2} - \frac{u(u-3s)}{t-m_G^2}\right] \nonumber\\
	&=& -\frac{ic_\ell^2}{4\Lambda^2} \left[\frac{(s+t)(s+4t)}{s-m_G^2}-\frac{(t+s)(t+4s)}{t-m_G^2}\right] \,.
\end{eqnarray}
Here the minus sign between the two terms in the bracket can be understood to come from the interchange the two external fermion particles of momenta $k_1$ and $p_2$.


\subsubsection{$\ell_R^-  \ell_L^+ \to \ell^-_L \ell_R^+$}
Now we consider another kinematic configuration $\ell_R^- \ell_L^+ \to \ell_L^- \ell_R^+$, in which the four momenta are denoted by 
\begin{eqnarray}
	p_1^\mu = (E,0,0,E)\,, &\quad& p_2^\nu = (E,0,0,-E)\,, \nonumber\\ 
	k_1^\rho = (E, E\sin\theta, 0, E\cos\theta)\,,  &\quad& k_2^\sigma = (E, -E\sin\theta, 0, -E\cos\theta)\,,
\end{eqnarray}
while their polarization spinors are given by
\begin{eqnarray}
	u_L(k_1) = \sqrt{2E} \left(\begin{array}{c}
		\xi_L \\
		0
	\end{array}\right) 
	=\sqrt{2E} \left(\begin{array}{c}
		-\sin\frac{\theta}{2}\\
		\cos\frac{\theta}{2}\\
		0\\
		0
	\end{array} \right)\,, \,
	v_R(k_2) =\sqrt{2E} \left(\begin{array}{c}
		\eta_L \\
		0
	\end{array}\right) = \sqrt{2E} \left(\begin{array}{c}
		\cos\frac{\theta}{2} \\
		\sin\frac{\theta}{2}\\
		0\\
		0
	\end{array}
	\right)\,.
\end{eqnarray}
For the $s$-channel, the amplitude after taking into account the on-shell condition can also be reduced to Eq.~(\ref{sllR}), so that by using the explicit expressions for momenta and polarization spinors of external particles, we have
\begin{eqnarray}
	\bar{v}_L(p_2) \gamma^\mu u_R(p_1) &=& (2E) (0,-1) \sigma^\mu \left(\begin{array}{c}
		1\\
		0
	\end{array}\right) = (2E) (0,-1,-i,0)^\mu\,,\nonumber\\
	\bar{u}_L(k_1)\gamma^\mu v_R(k_2) &=& (2E) (-\sin\frac{\theta}{2}, \cos\frac{\theta}{2}) \bar{\sigma}^\mu \left(\begin{array}{c}
		\cos\frac{\theta}{2} \\
		\sin\frac{\theta}{2} 
	\end{array}\right) = (0,-\cos\theta, -i, \sin\theta)^\mu\,,
\end{eqnarray}
and
\begin{eqnarray}
	\left[ \bar{u}_L(k_1) \gamma^\mu v_R(k_2) \right] \left[\bar{v}_L(p_2) \gamma_\mu u_R(p_1)\right] &=& (2E)^2 (1-\cos\theta) = -2t\,,\nonumber\\
	\bar{u}_L(k_1) (\slashed{p}_1-\slashed{p}_2) v_R(k_2) &=& -(2E)^2 \sin\theta\,,\nonumber\\
	\bar{v}_L(p_2) (\slashed{k}_1-\slashed{k}_2) u_R(p_1) &=& (2E)^2 \sin\theta\,.
\end{eqnarray}
Therefore, the $s$-channel amplitude is given by
\begin{eqnarray}
	i{\cal M}_s(\ell_R^- \ell_L^+ \to \ell_L^- \ell_R^+) = -\frac{ic_\ell^2}{4\Lambda^2} \frac{t(t-3u)}{s-m_G^2}\,.
\end{eqnarray}
Note that the bi-spinor forms like $\bar{u}_L(k_1)\gamma^\mu u_R(p_1)=0$ or $\bar{v}_L(p_2) \gamma^\mu v_R(k_2) = 0$ vanish, so that we cannot write down the $t$-channel amplitude. Thus, the total amplitude of $\ell_R^- \ell_L^+ \to \ell_L^- \ell_R^+$ is given by its $s$-channel one, with the following final result
\begin{eqnarray}
	i{\cal M}(\ell_R^- \ell_L^+ \to \ell_L^- \ell_R^+) = 	i{\cal M}_s(\ell_R^- \ell_L^+ \to \ell_L^- \ell_R^+)  = -\frac{ic_\ell^2}{4\Lambda^2} \frac{t(t-3u)}{s-m_G^2}\,.
\end{eqnarray}

\subsubsection{$\ell_R^- \ell_R^+ \to \ell_R^- \ell_R^+$}
Since we cannot write the bi-spinors, such as $\bar{v}_R(p_2) \gamma^\mu u_R(p_1)$ or $\bar{u}_R(k_1) \gamma^\mu v_R(k_2)$, the $s$-channel diagram cannot provide any contribution to this kinematic configuration. However, the $t$-channel Feynman diagram does contribute, and in the following we would like to compute it. By using the on-shell conditions, we can simplify the $t$-channel amplitude into the following form
\begin{eqnarray}
	i{\cal M}_t(\ell_R^- \ell_R^+ \to \ell_R^- \ell_R^+) =&& \frac{i c_\ell^2}{8\Lambda^2} \frac{1}{q^2 - m_G^2} \bigg\{ (k_1+p_1) (k_2+p_2) [\bar{u}_R(k_1) \gamma^\mu u_R(k_2)][\bar{v}_R(p_2) \gamma_\mu v_R(k_2)] \nonumber\\
	&+& [\bar{u}_R(k_1)(\slashed{p}_2 + \slashed{k}_2) u_R(p_1)][\bar{v}_R(p_2) (\slashed{p}_1+\slashed{k}_1)v_R(k_2)]	\bigg\}\,.
\end{eqnarray}
Here by taking the explicit form of momenta and spinors in the com frame, we can obtain
\begin{eqnarray}
	\bar{u}_R(k_1) \gamma^\mu u_R(p_1) &=& (2E) (\cos\frac{\theta}{2}, \sin\frac{\theta}{2}) \sigma^\mu \left(\begin{array}{c}
		1 \\
		0
	\end{array}\right) = (2E) \left(\cos\frac{\theta}{2}, \sin\frac{\theta}{2}, i\sin\frac{\theta}{2}, \cos\frac{\theta}{2} \right)^\mu\,, \nonumber\\
	\bar{v}_R(p_2) \gamma^\mu v_R(k_2) &=&  (2E) (1,0) \bar{\sigma}^\mu \left(\begin{array}{c}
		\cos\frac{\theta}{2} \\
		\sin\frac{\theta}{2}
	\end{array}\right) = (2E) \left(\cos\frac{\theta}{2}, -\sin\frac{\theta}{2}, i\sin\frac{\theta}{2}, -\cos\frac{\theta}{2}\right)^\mu\,.
\end{eqnarray}
so that
\begin{eqnarray}
	[\bar{u}_R(k_1) \gamma^\mu u_R(p_1)][\bar{v}_R(p_2) \gamma_\mu v_R(k_2)] &=& 2(2E)^2 = 2s\,, \nonumber\\
	\bar{u}_R(k_1) (\slashed{p}_2 + \slashed{k}_2) u_R(p_1) &=& 2 (2E)^2 \cos\frac{\theta}{2}\,, \nonumber\\
	\bar{v}_R(p_2) (\slashed{p}_1 + \slashed{k}_1) v_R(k_2) &=&  2 (2E)^2 \cos\frac{\theta}{2}\,.
\end{eqnarray}
It turns out that the final expression for the amplitude of $\ell_R^- \ell_R^+ \to \ell_R^- \ell_R^+$ is given by
\begin{eqnarray}
	i{\cal M}(\ell_R^- \ell_R^+ \to \ell_R^- \ell_R^+) = i{\cal M}_t(\ell_R^- \ell_R^+ \to \ell_R^- \ell_R^+) = \frac{ic_\ell^2}{4\Lambda^2} \frac{s(s-3u)}{t-m_G^2}\,.
\end{eqnarray}

\subsubsection{Unitarity Bounds for $\ell^- \ell^+ \to \ell^- \ell^+$}
\begin{figure}[!htb]
	\centering
	\hspace{-5mm}
	\includegraphics[width=0.7\linewidth]{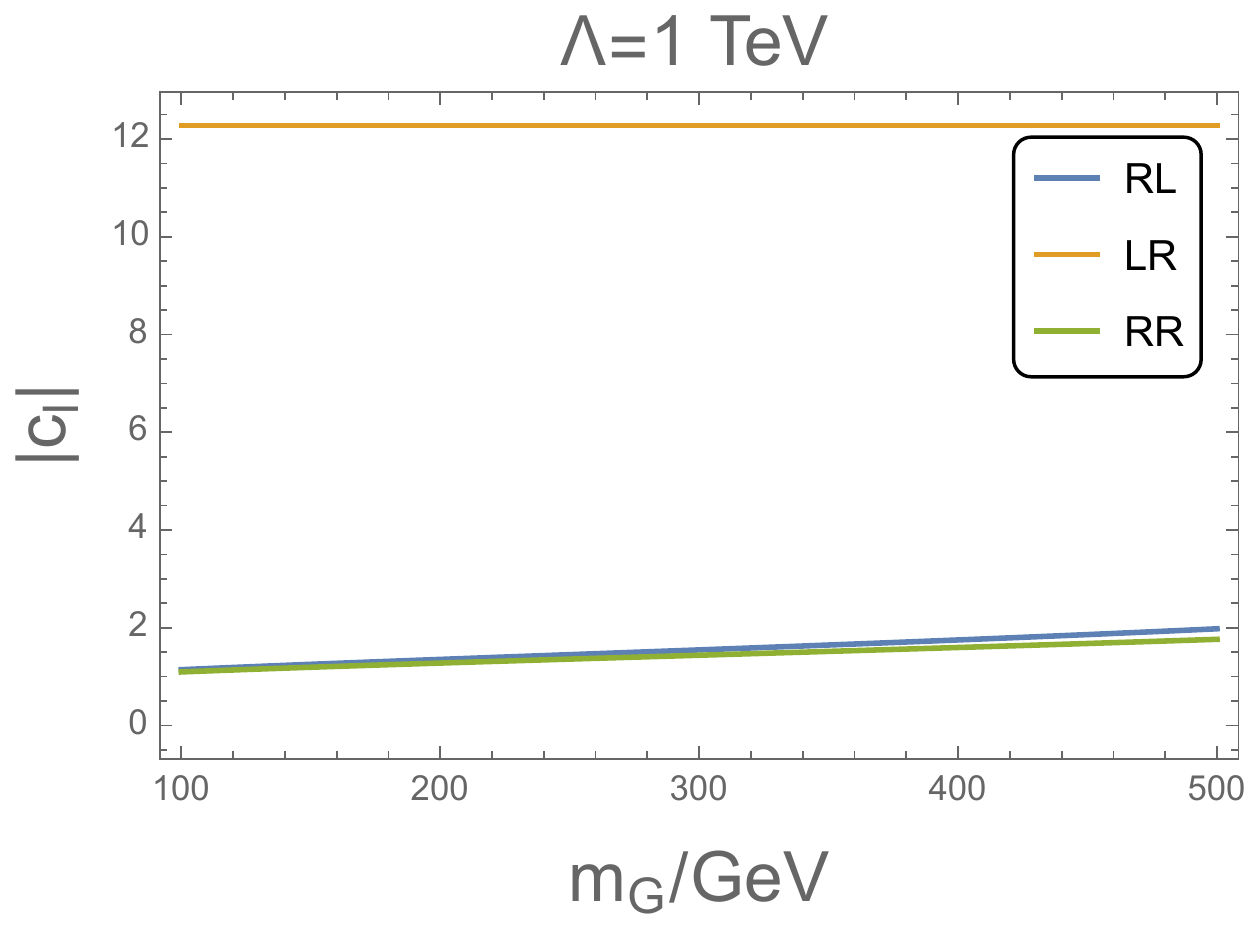}
	\caption{The unitarity bounds on the lepton-$G$ Wilsonian coefficients $c_\ell$  for the three independent helicity configurations: $\ell^-_R \ell^+_L \to \ell^-_R \ell^+_L$, $\ell^-_R \ell^+_L \to \ell^-_L \ell^+_R$ and $\ell^-_R \ell^+_R \to \ell^-_R \ell^+_R$, which are labeled as RL, LR, and RR in the legend. } \label{UniBoundLep}
\end{figure}

In the above discussion, we obtains all of the nonzero $\ell^-\ell^+ \to \ell^- \ell^+$ amplitudes of different helicity configurations, which are summarized again as follows
\begin{eqnarray}
	{\cal M} (\ell^-_R \ell^+_L \to \ell^-_R \ell^+_L) &=& {\cal M} (\ell^-_L \ell^+_R \to \ell^-_L \ell^+_R) = -\frac{c_\ell^2}{4\Lambda^2}\left[ \frac{u(u-3t)}{s-m_G^2} - \frac{u(u-3s)}{t-m_G^2} \right]\,,\nonumber\\
	{\cal M}	 (\ell^-_R \ell^+_R \to \ell^-_R \ell^+_R) &=& {\cal M} (\ell^-_L \ell^+_L \to \ell^-_L \ell^+_L) = \frac{c_\ell^2}{4\Lambda^2} \frac{s(s-3u)}{t-m_G^2}\,, \nonumber\\
	{\cal M} (\ell^-_R \ell^+_L \to \ell^-_L \ell^+_R) &=& {\cal M} (\ell^-_L \ell^+_R \to \ell^-_R \ell^+_L) = -\frac{c_\ell^2}{4\Lambda^2} \frac{t(t-3u)}{s-m_G^2}\,,
\end{eqnarray} 
where the two amplitudes in a single line are $P$-symmetric to one another, while other helicity configurations would lead to vanishing amplitudes. By using the definition of $a_0(\sqrt{s})$ in Eq.~(\ref{DefA0}), we can obtain the $s$-wave projected amplitudes, given by
\begin{eqnarray}
	&& a_0(\ell^-_R \ell^+_L \to \ell^-_R \ell^+_L) = a_0(\ell^-_L \ell^+_R \to \ell^-_L \ell^+_R) \nonumber\\
	&=& -\frac{1}{16\pi s} \frac{c_\ell^2}{4\Lambda^2} \left[-\frac{s(28s^2 - 21 m_G^2 s -6 m_G^4)}{6(s-m_G^2)} + (4s+m_G^2) (s+m_G^2) \ln\left(\frac{s+m_G^2}{m_G^2}\right)  \right] \nonumber\\
	&\approx & \frac{c_\ell^2}{16\pi} \frac{s}{4\Lambda^2} \left( \frac{14}{3} - 4\ln\frac{s}{m_G^2} \right) \sim \frac{c_\ell^2}{16\pi} \left(\frac{14}{3} - 4\ln \frac{4\Lambda^2}{m_G^2}\right)\,,\nonumber\\
	&& a_0 (\ell^-_R \ell^+_R \to \ell^-_R \ell^+_R) = a_0 (\ell^-_L \ell^+_L \to \ell^-_L \ell^+_L) \nonumber\\
	&=& \frac{1}{16\pi} \frac{c_\ell^2}{4\Lambda^2} \left[ 3s -(4s +3m_G^2) \ln \left( \frac{s+m_G^2}{m_G^2} \right) \right] \approx \frac{c_\ell^2}{16\pi} \frac{s}{4\Lambda^2} \left( 3 - 4\ln\frac{s}{m_G^2} \right) \nonumber\\
	&\sim& \frac{c_\ell^2}{16\pi} \left( 3-4\ln\frac{4\Lambda^2}{m_G^2} \right)\,, \nonumber\\
	&& a_0 (\ell^-_R \ell^+_L \to \ell^-_L \ell^+_R) = a_0 (\ell^-_L \ell^+_R\to \ell^-_R \ell^+_L) \nonumber\\
	&=& \frac{1}{16\pi}\frac{c_\ell^2}{4\Lambda^2}\frac{s^2}{6(s-m_G^2)} \approx  \frac{1}{16\pi}\frac{c_\ell^2}{6} \frac{s}{4\Lambda^2} \sim \frac{c_\ell^2}{96\pi}\,.
\end{eqnarray}
where the symbol $\sim$ denotes the high energy limit with $s \to 4\Lambda^2$. By requiring the $s$-wave projected amplitudes to satisfy the unitarity bound ${\rm Re} [a_0(\sqrt{s})] \leq 1/2 $, we can obtain the following constraints on the Wilsonian coefficient $c_\ell$
\begin{eqnarray}
	\ell^-_R \ell^+_L \to \ell^-_R \ell^+_L : \quad	c_\ell &\leq& \sqrt{\frac{8\pi}{\left| \frac{14}{3} - 4\ln\frac{4\Lambda^2}{m_G^2} \right|}}\,,\\
	\ell^-_R \ell^+_L \to \ell^-_L \ell^+_R : \quad		c_\ell &\leq& \sqrt{48\pi}\,,\\
	\ell^-_R \ell^+_R \to \ell^-_R \ell^+_R : \quad
	c_\ell &\leq & \sqrt{\frac{8\pi}{\left| 3 - 4\ln\frac{4\Lambda^2}{m_G^2} \right|}}\,. \label{UBLepton}
\end{eqnarray}
In Fig.~\ref{UniBoundLep}, we plot the upper limits on $c_\ell$ derived from different helicity states, which are functions of the spin-2 particle mass $m_G$ when $\Lambda = 1$~TeV. It is seen that the most stringent constraint on $c_\ell$ is given by the channel $\ell^-_R \ell^+_R \to \ell^-_R \ell^+_R$ ($\ell^-_L \ell^+_L \to \ell^-_L \ell^+_L$) in the spin-2 particle mass range from $100$~GeV to 500~GeV, which is of great interest to in the interpretation of the muon $g-2$ anomaly.

\subsection{$\gamma\gamma\to \gamma\gamma$}
\begin{figure}[!htb]
	\centering
	\hspace{-5mm}
	\includegraphics[width=0.32\linewidth]{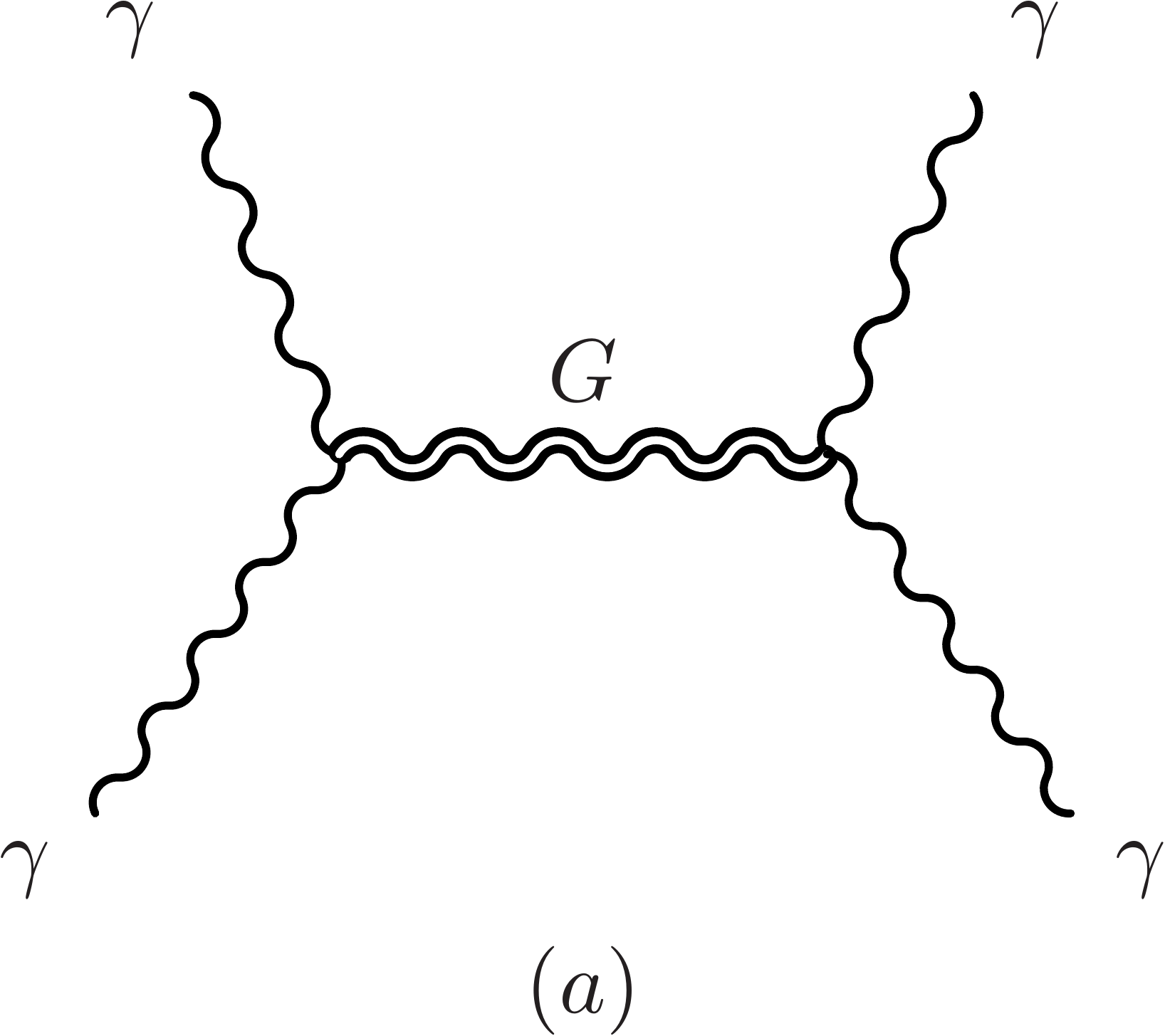}
	\includegraphics[width=0.32\linewidth]{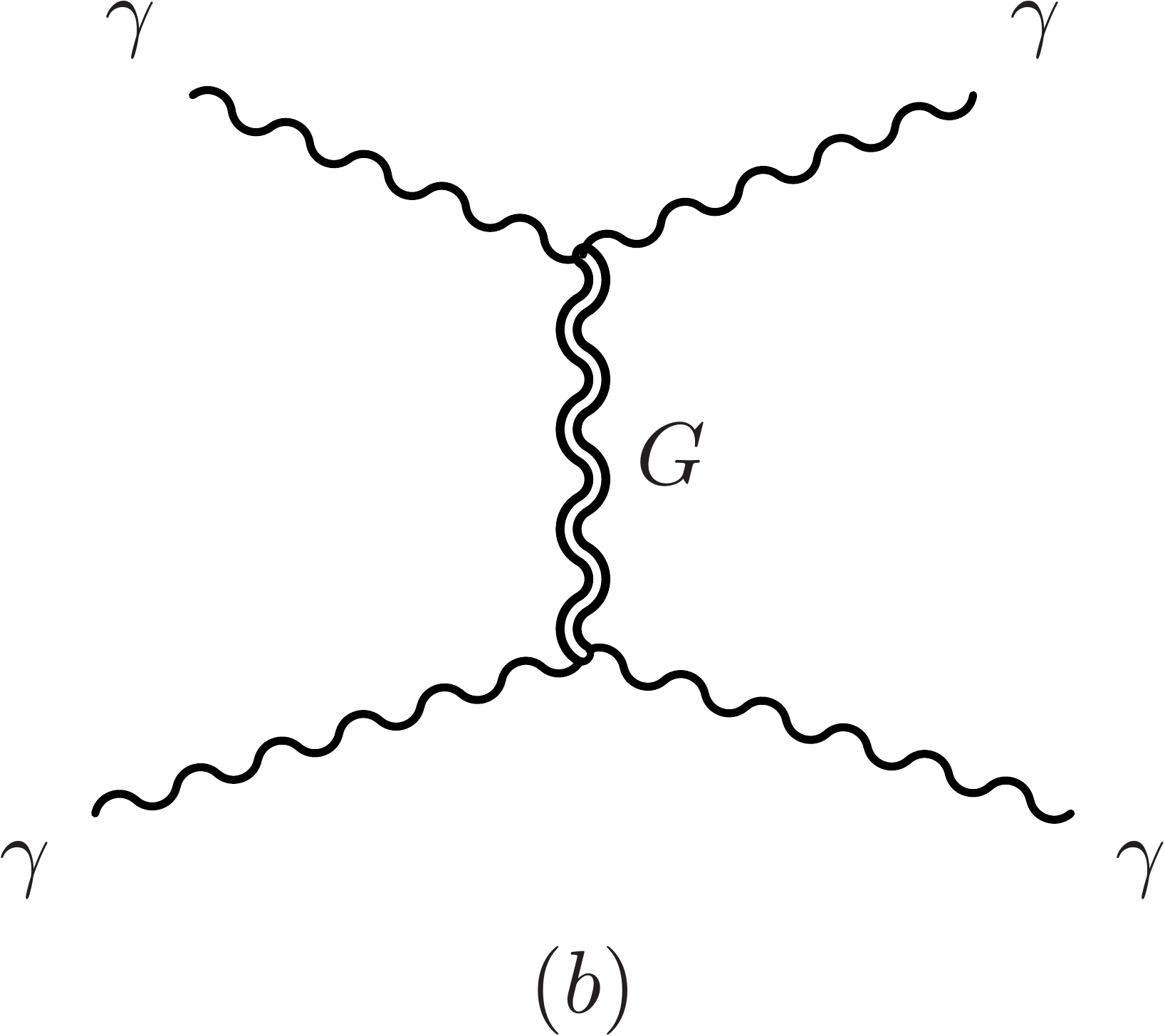}
	\includegraphics[width=0.32\linewidth]{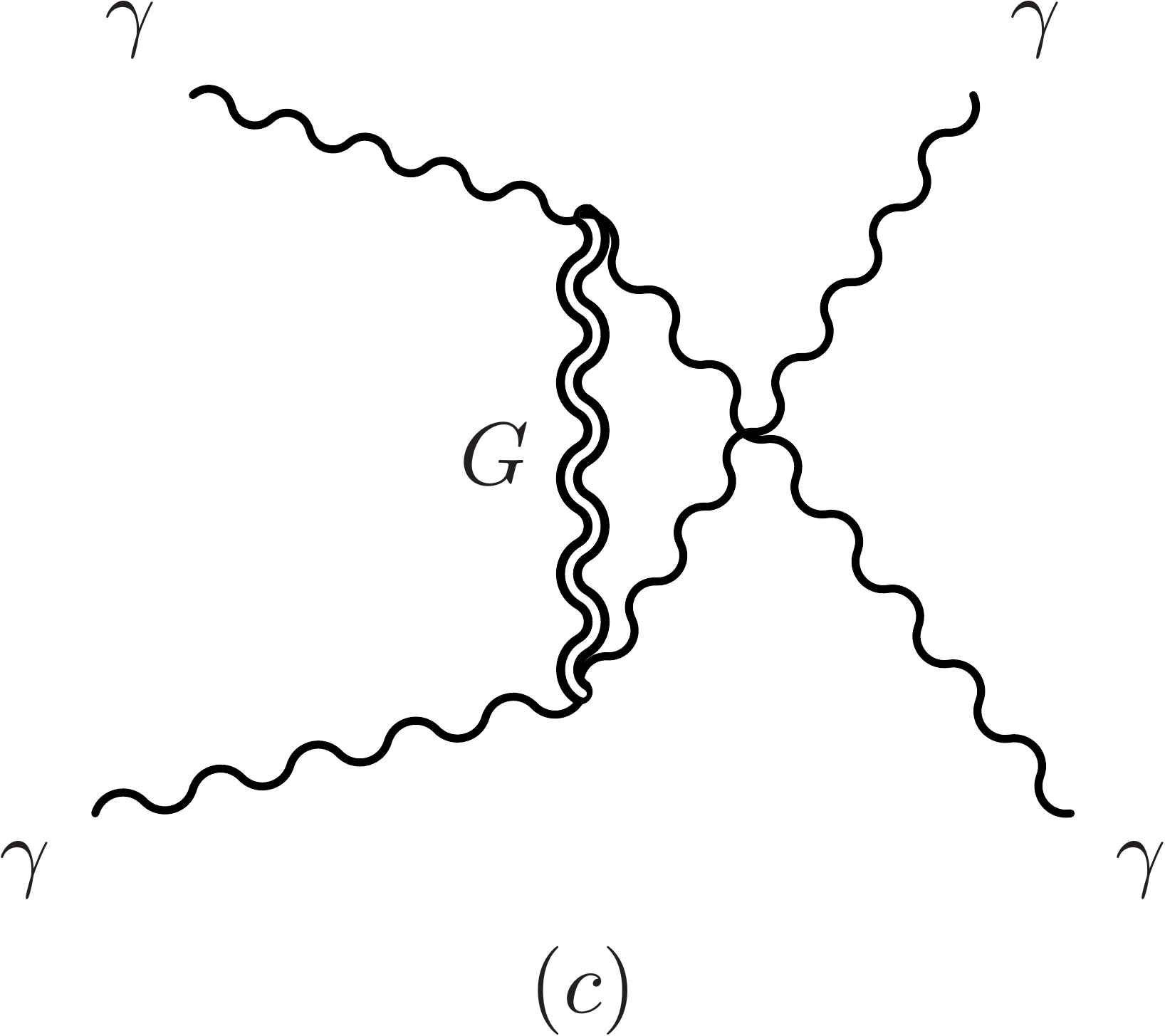}
	\caption{Feynman diagrams for the photon-photon scattering process of $\gamma\gamma\to \gamma\gamma$.}\label{Figgg}
\end{figure}

In order to obtain the unitarity bounds on the photon-$G$ coupling $c_\gamma$, we have to calculate the $\gamma\gamma \to \gamma\gamma$ amplitudes of various polarization configurations. The relevant Feynman diagrams are shown in Fig.~\ref{Figgg}, which correspond to the $s$-, $t$-, and $u$-channels, respectively. As the photon-$G$ vertex in Eq.~(\ref{LagG}) preserves the spatial parity $P$ symmetry, the amplitudes of different photon polarizations are the same to each other when they are related by parity transformation, which greatly reduces the number of independent photon 2-to-2 scattering amplitudes. 
In this subsection, we compute these independent nonzero photon scattering amplitudes, based on which we can yield the associated unitarity bounds on $c_\gamma$.

\subsubsection{$\gamma_R\gamma_R \to \gamma_R \gamma_R$}
Firstly, we consider the amplitude of $\gamma_R \gamma_R \to \gamma_R \gamma_R$, where the four external photons are all right-handed.
In the com frame, the two incoming photons are specified by
\begin{eqnarray}\label{gIn}
	p_1^\mu = (E, 0, 0, E)\,, &\quad& \epsilon^\mu_R (p_1) = \frac{1}{\sqrt{2}} (0,-1,-i, 0)\,, \nonumber\\
	p_2^\nu = (E,0,0,-E)\,, &\quad& \epsilon^\nu_R(p_2) = \frac{1}{\sqrt{2}} (0,1, -i, 0)\,,
\end{eqnarray}
while the two outgoing photons are represented by
\begin{eqnarray}\label{gOut}
	k_1^\rho = (E, E\sin\theta, 0 , E\cos\theta)\,, &\quad& \epsilon^*_R(k_1) = \frac{1}{\sqrt{2}} (0, -\cos\theta, i, \sin\theta)\,, \nonumber\\
	k_2^\sigma = (E, -E\sin\theta, 0, -E\cos\theta)\,, &\quad& \epsilon^*_R(k_2) = \frac{1}{\sqrt{2}} (0,\cos\theta, i, -\sin\theta)\,.
\end{eqnarray}
The amplitude constitutes three Feynman diagrams shown in Fig.~\ref{Figgg}, corresponding to $s$-, $t$- and $u$-channels, respectively. In the following, we shall calculate them one by one. According to the Feynman rules, the $s$-channel amplitude is given by
\begin{eqnarray}
	i{\cal M}_s (\gamma_R \gamma_R \to \gamma_R \gamma_R) &=& -\frac{ic_\gamma^2}{2\Lambda^2} \epsilon_R^\mu (p_1) \epsilon_R^\nu(p_2) \epsilon_R^*(k_1) \epsilon^*_R (k_2) [(p_1\cdot p_2) C_{\alpha\beta, \mu\nu} + D_{\alpha\beta, \mu\nu}(p_1, p_2)] \nonumber\\
	&& \frac{B^{\alpha \beta, \lambda\delta}(Q)}{Q^2-m_G^2}  [(k_1\cdot k_2) C_{\lambda\delta, \rho\sigma} + D_{\lambda\delta, \rho\sigma}(k_1, k_2)]\,.
\end{eqnarray}
Note that, by using the relation $Q= p_1 +p_2 = k_1 + k_2$, we find that terms proportional to four- and two-powers of $1/m_G$ are canceled. By further making use of the specific momenta and polarization vectors of the four external particles in Eqs.~(\ref{gIn}) and (\ref{gOut}), the $s$-channel amplitude vanishes, while the $t$- and $u$-channel amplitudes are given by
\begin{eqnarray}
	i{\cal M}_t (\gamma_R \gamma_R \to \gamma_R \gamma_R) &=& -\frac{ic_\gamma^2}{2\Lambda^2} \epsilon_R^\mu (p_1) \epsilon_R^\nu(p_2) \epsilon_R^*(k_1) \epsilon^*_R (k_2) [(-p_1\cdot k_1) C_{\alpha\beta, \mu\rho} + D_{\alpha\beta, \mu\rho}(p_1, -k_1)] \nonumber\\
	&& \frac{B^{\alpha \beta, \lambda\delta}(q)}{q^2-m_G^2}  [(-p_2\cdot k_2) C_{\lambda\delta, \nu\sigma} + D_{\lambda\delta, \nu\sigma}(p_2, -k_2)] \nonumber\\
	&=& -\frac{ic_\gamma^2}{2\Lambda^2} \frac{2s^2}{t-m_G^2}\,, \\
	i{\cal M}_u (\gamma_R \gamma_R \to \gamma_R \gamma_R) &=& -\frac{ic_\gamma^2}{2\Lambda^2} \epsilon_R^\mu (p_1) \epsilon_R^\nu(p_2) \epsilon_R^*(k_1) \epsilon^*_R (k_2) [(-p_1\cdot k_2) c_{\alpha\beta, \mu\rho} + D_{\alpha\beta, \mu\rho}(p_1, -k_2)] \nonumber\\
	&& \frac{B^{\alpha \beta, \lambda\delta}(q^\prime)}{q^{\prime \, 2}-m_G^2}  [(-p_2\cdot k_1) C_{\lambda\delta, \nu\sigma} + D_{\lambda\delta, \nu\sigma}(p_2, -k_1)] \nonumber\\
	&=& -\frac{ic_\gamma^2}{2\Lambda^2} \frac{2s^2}{u-m_G^2}\,,
\end{eqnarray} 
where $q\equiv p_1 - k_1 = k_2 -p_2$ and $q^\prime = p_1 - k_2 = k_1 -p_2$, so that $q^2 = t$ and $q^{\prime \, 2} = u$. Therefore, the total amplitude of $\gamma_R \gamma_R \to \gamma_R \gamma_R$ is written as 
\begin{eqnarray}\label{gg0}
	i{\cal M}(\gamma_R \gamma_R \to \gamma_R \gamma_R) = -\frac{iC_\gamma^2}{2\Lambda^2} \left(\frac{2s^2}{t-m_G^2}+ \frac{2s^2}{u-m_G^2}\right)\,.
\end{eqnarray}

\subsubsection{$\gamma_R\gamma_L \to \gamma_R \gamma_L$}
We now turn to the process of $\gamma_R \gamma_L \to \gamma_R \gamma_L$. In the com frame, we can define the polarization vectors of the incoming particles as follows
\begin{eqnarray}
	\epsilon_R^\mu(p_1) = \frac{1}{\sqrt{2}} (0,-1,-i,0)^\mu\,, \quad \epsilon_L^\nu = \frac{1}{\sqrt{2}} (0,-1,-i,0)^\nu\,, 
\end{eqnarray}
while those of the outgoing particles are given by
\begin{eqnarray}
	\epsilon_R^* (k_1) = \frac{1}{\sqrt{2}} (0,-\cos\theta, i, \sin\theta)\,, \quad \epsilon^*_L (k_2) = \frac{1}{\sqrt{2}} (0, -\cos\theta, i, \sin\theta)\,.
\end{eqnarray}
By taking the above expressions of the polarization vectors and momenta of the external particles into formulas for the $s$-, $t$- and $u$-channels, we can easily obtain the following partial amplitudes
\begin{eqnarray}
	i{\cal M}_s (\gamma_R \gamma_L \to \gamma_R \gamma_L) &=& -\frac{ic_\gamma^2}{2\Lambda^2} \frac{2u^2}{s-m_G^2}\,, \nonumber\\
	i{\cal M}_t (\gamma_R \gamma_L \to \gamma_R \gamma_L) &=& -\frac{ic_\gamma^2}{2\Lambda^2} \frac{2u^2}{t-m_G^2}\,, \nonumber\\
	i{\cal M}_u (\gamma_R \gamma_L \to \gamma_R \gamma_L) &=& 0\,.
\end{eqnarray}
Thus, the total amplitude is the summation over the $s$-, $t$- and $u$-channels
\begin{eqnarray}
	i{\cal M} (\gamma_R \gamma_L \to \gamma_R \gamma_L) = -\frac{ic_\gamma^2}{2\Lambda^2} \left(\frac{2u^2}{s-m_G^2} + \frac{2u^2}{t-m_G^2}\right)\,.
\end{eqnarray}

\subsubsection{$\gamma_R \gamma_L \to \gamma_L \gamma_R$}
In the case of $\gamma_R \gamma_L \to \gamma_L \gamma_R$, the polarization vectors for the external photons in the com frame are given by
\begin{eqnarray}
	&& \epsilon_R (p_1) = \frac{1}{\sqrt{2}} (0, -1, -i, 0)\,,\quad \epsilon_L (p_2) = \frac{1}{\sqrt{2}} (0, -1, -i, 0) \,, \nonumber\\
	&& \epsilon^*_L(k_1) = \frac{1}{\sqrt{2}} (0, \cos\theta, i -\sin\theta)\,, \quad \epsilon_R^* (k_2) = \frac{1}{\sqrt{2}} (0, \cos\theta, i, -\sin\theta)\,.
\end{eqnarray}
Following exactly the same procedure in the previous subsections, we can derive the following amplitudes for the $s$-, $t$- and $u$-channels of $\gamma_R \gamma_L \to \gamma_L \gamma_R$:
\begin{eqnarray}
	i{\cal M}_s (\gamma_R \gamma_L \to \gamma_L \gamma_R) &=& -\frac{ic_\gamma^2}{2\Lambda^2} \frac{2t^2}{s-m_G^2}\,, \nonumber\\
	i{\cal M}_t (\gamma_R \gamma_L \to \gamma_L \gamma_R) &=& 0 \,, \nonumber\\
	i{\cal M}_u (\gamma_R \gamma_L \to \gamma_L \gamma_R) &=& -\frac{ic_\gamma^2}{2\Lambda^2} \frac{2t^2}{u-m_G^2}\,. 
\end{eqnarray}
It turns out that the total amplitude of $\gamma_R \gamma_L \to \gamma_L \gamma_R$ is given by
\begin{eqnarray}
	i{\cal M} (\gamma_R \gamma_L \to \gamma_L \gamma_R) = -\frac{ic_\gamma^2}{2\Lambda^2} \left( \frac{2t^2}{s-m_G^2} + \frac{2t^2}{u-m_G^2} \right)\,.
\end{eqnarray}

\subsubsection{Unitarity Bounds for $\gamma\gamma \to \gamma\gamma$}
Here we summarize the nonzero independent $\gamma\gamma\to \gamma\gamma$ amplitudes of different polarizations as follows
\begin{eqnarray}
	i{\cal M} (\gamma_R \gamma_L \to \gamma_R \gamma_L) &=& i{\cal M} (\gamma_L\gamma_R \to \gamma_L \gamma_R) = -\frac{ic_\gamma^2}{2\Lambda^2} \left( \frac{2u^2}{s-m_G^2} + \frac{2u^2}{t-m_G^2} \right)\,,\nonumber\\
	i{\cal M}  (\gamma_R \gamma_L \to \gamma_L \gamma_R) &=& i{\cal M} (\gamma_L \gamma_R \to \gamma_R \gamma_L)  =  -\frac{ic_\gamma^2}{2\Lambda^2} \left( \frac{2t^2}{s-m_G^2} + \frac{2t^2}{u-m_G^2} \right)\,,\nonumber\\
	i{\cal M}  (\gamma_R \gamma_R \to \gamma_R \gamma_R) &=& i{\cal M} (\gamma_L \gamma_L \to \gamma_L \gamma_L)  =  -\frac{ic_\gamma^2}{2\Lambda^2} \left( \frac{2s^2}{t-m_G^2} + \frac{2s^2}{u-m_G^2} \right)\,,
\end{eqnarray} 
while all other polarization amplitudes vanish identically. According to the definition of $a_0(\sqrt{s})$ in Eq.~(\ref{DefA0}), the $s$-wave projected amplitudes for the nontrivial polarization assignments are given by
\begin{eqnarray}
	&& a_0(\gamma_R \gamma_R \to \gamma_R \gamma_R) = a_0 (\gamma_L\gamma_L \to \gamma_L\gamma_L) \nonumber\\
	&=& \frac{c_\gamma^2 }{4\pi} \frac{s}{4\Lambda^2} \ln \frac{s+m_G^2}{m_G^2} \approx \frac{c_\gamma^2 }{4\pi} \frac{s}{4\Lambda^2} \ln \frac{s}{m_G^2} 
	\sim  \frac{c_\gamma^2}{4\pi} \ln \frac{4\Lambda^2}{m_G^2}\,, \nonumber\\
	&& a_0(\gamma_R \gamma_L \to \gamma_R \gamma_L) = a_0(\gamma_R \gamma_L \to \gamma_L \gamma_R) = a_0(\gamma_L \gamma_R \to \gamma_L \gamma_R) = a_0(\gamma_L \gamma_R \to \gamma_R \gamma_L) \nonumber\\
	&=& -\frac{1}{32\pi s}\frac{c_\gamma^2}{\Lambda^2} \left[ \frac{s(11s^2 - 3 m_G^2 s-6m_G^4)}{6(s-m_G^2)} -(s+m_G^2) \ln\frac{s+m_G^2}{m_G^2} \right] \nonumber\\
	&\approx & -\frac{c_\gamma^2}{8\pi}\frac{s}{4\Lambda^2} \left(\frac{11}{6} - \ln\frac{s}{m_G^2}\right) \sim -\frac{c_\gamma^2}{8\pi} \left( \frac{11}{6} - \ln\frac{4\Lambda^2}{m_G^2} \right)\,,
\end{eqnarray}
where we have taken the high-energy limit with $s\sim 4\Lambda^2 \gg m_G^2$. 
Therefore, the unitarity bounds for these channels are given by
\begin{eqnarray}
	&& \gamma_R \gamma_R \to \gamma_R \gamma_R: \quad c_\gamma \lesssim \sqrt{\frac{2\pi}{\left|\ln\frac{4\Lambda^2}{m_G^2}\right|}}\,, \label{UBGamma} \\
	&& \gamma_R \gamma_L \to \gamma_R \gamma_L: \quad  c_\gamma \lesssim \sqrt{\frac{4\pi}{\left| \frac{11}{6} - \ln\frac{4\Lambda^2}{m_G^2} \right|}}\,.
\end{eqnarray}

\begin{figure}[!htb]
	\centering
	\hspace{-5mm}
	\includegraphics[width=0.7\linewidth]{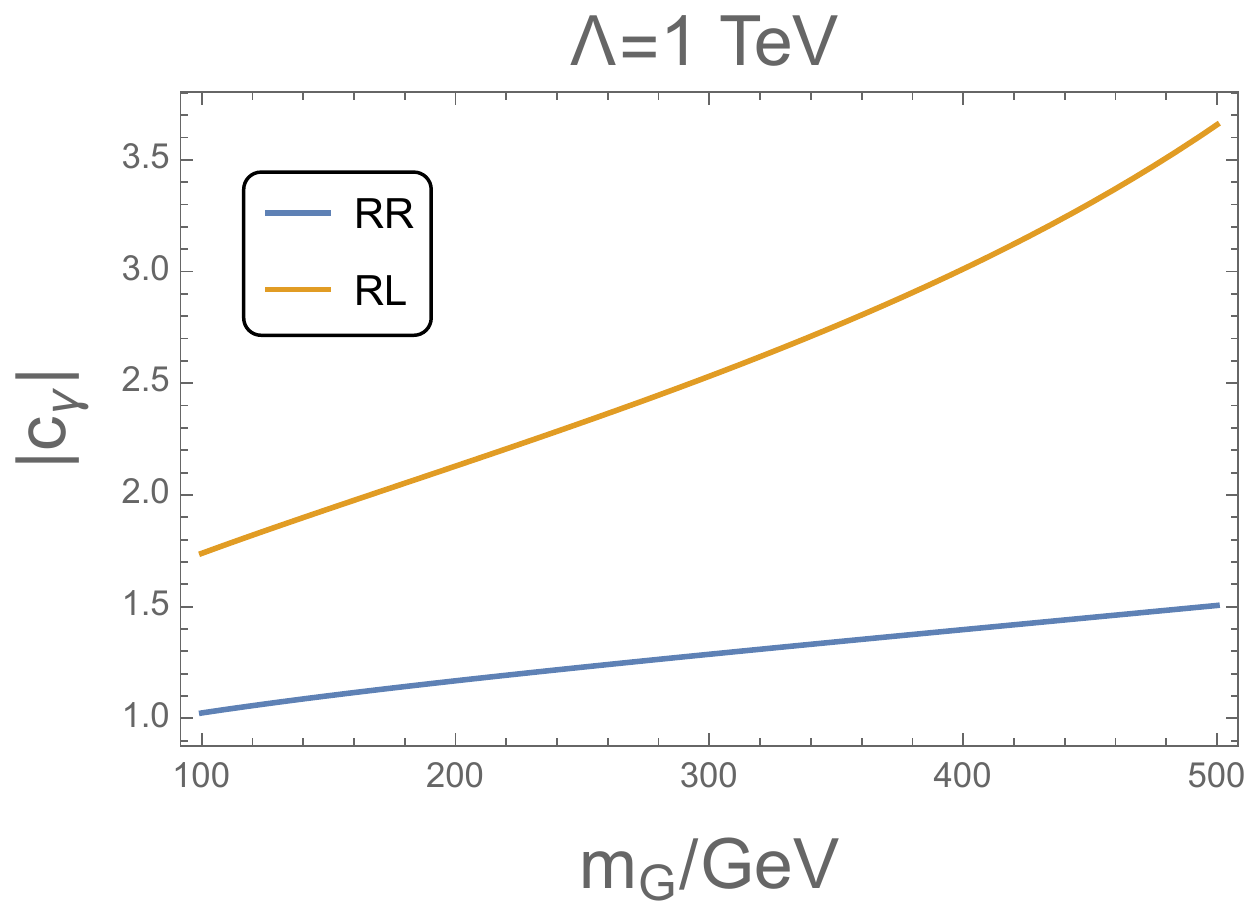}	
	\caption{The unitarity bounds on the photon-$G$ Wilsonian coefficient $c_\gamma$ for two independent helicity configurations: $\gamma_R \gamma_R \to \gamma_R \gamma_R$ and $\gamma_R \gamma_L \to \gamma_R \gamma_L$, which are labeled as RR and RL in the legend.  } \label{UniBoundGam}
\end{figure}
In Fig.~\ref{UniBoundGam}, we make plots for these unitarity bounds, which are set on the coupling $c_\gamma$ as a function of the spin-2 particle mass $m_G$ with a fixed cutoff scale $\Lambda = 1$~TeV. It is clear that the constraint from the channel $\gamma_R \gamma_R \to \gamma_R \gamma_R$ is much stronger than that from $\gamma_R\gamma_L\to \gamma_R \gamma_L$ for $m_G>100$~GeV. Hence, we will apply the upper limit from $\gamma_R \gamma_R \to \gamma_R \gamma_R$ as our unitarity bound in the following numerical analysis.

\section{Numerical Studies}\label{SecResult}
In this section, we study the constraints on the model parameter space by exploiting the unitarity bounds in Eqs.~(\ref{UBLepton}) and (\ref{UBGamma}), which sets upper limits on the massive spin-2 particle couplings $c_\ell$ and $c_\gamma$. Note that the spin-2 particle model in Ref.~\cite{Huang:2022} has been proposed to explain the long-standing lepton $g-2$ anomaly. Currently, the most precise value of the muon $g-2$ is provided by the combined data from Brookhaven and Fermilab, which deviates the SM predictions by 4.25 $\sigma$~CL. On the other hand, the measurements of the fine structure constant $\alpha$ at Laboratoire Kastler Brossel (LKB)~\cite{Morel:2020dww} and  Berkeley~\cite{Parker:2018vye} have given two latest SM predictions~\cite{Aoyama:2012wj,Aoyama:2019ryr} on the electron anomalous magnetic dipole moment, $a_e^{\rm LKB}$ and $a_e^{\rm B}$, with their differences from the experimental result $a^{\rm exp}_e$~\cite{Hanneke:2008tm} given by
\begin{eqnarray}\label{g2Electron}
	\Delta a_e^{\rm LKB} &=& a_e^{\rm exp} - a_e^{\rm LKB} = (4.8\pm 3.0) \times 10^{-13}\,, \nonumber\\
	\Delta a_e^{\rm B} &=& a_e^{\rm exp} - a_e^{\rm B} = (-8.8 \pm 3.6) \times 10^{-13}\,.  
\end{eqnarray}   
Based on the general lepton $g-2$ formula in Eq.~(\ref{G2total}), we can also discuss the implications of our spin-2 particle model on the electron $g-2$. Since the data in Eq.~(\ref{g2Electron}) given at LKB and  Berkeley are incompatible with each other, we would like to discuss them separately.  

\begin{figure}[!htb]
	\centering
	\hspace{-5mm}
	\includegraphics[width=0.5\linewidth]{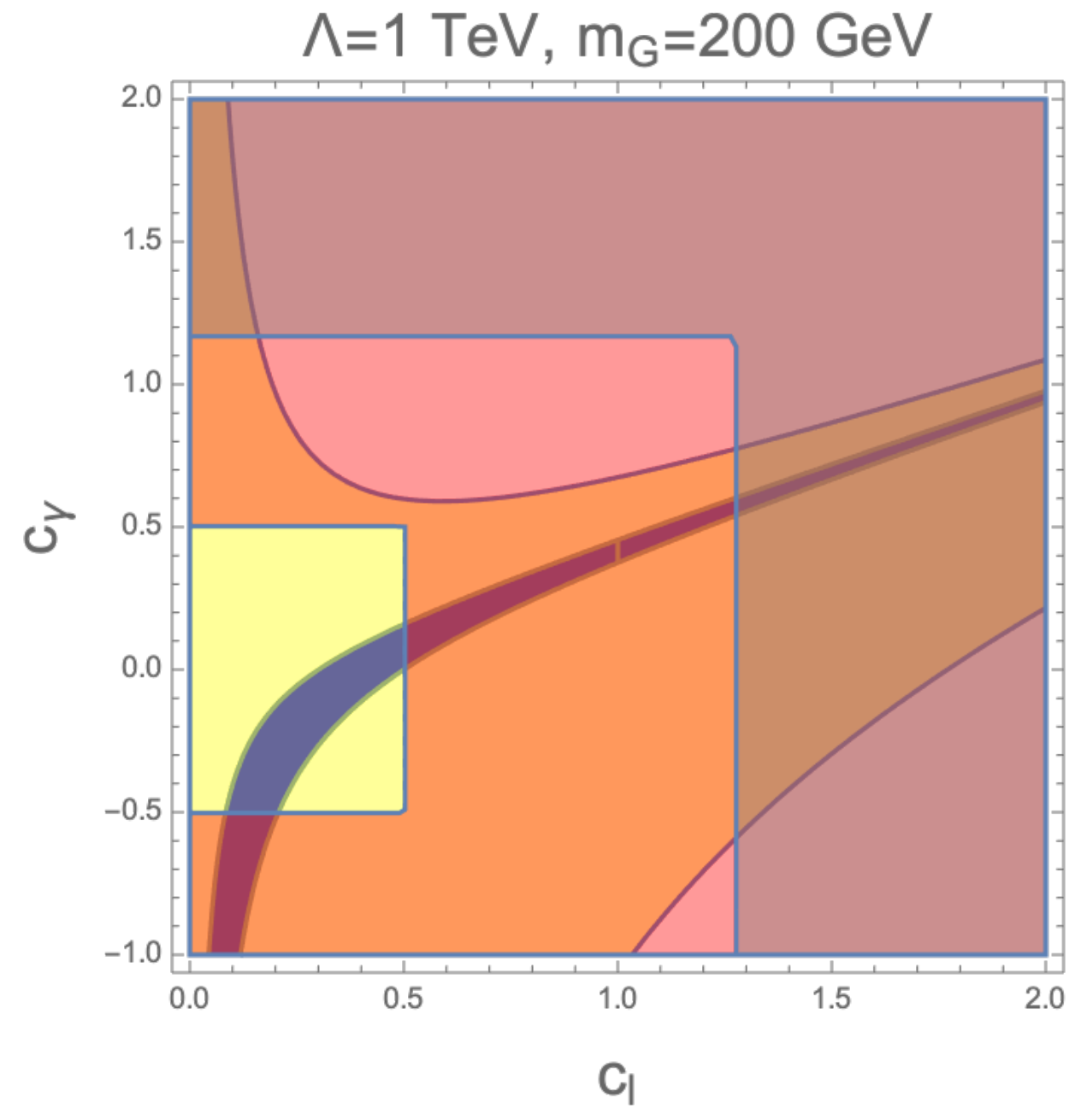}
	\includegraphics[width=0.5\linewidth]{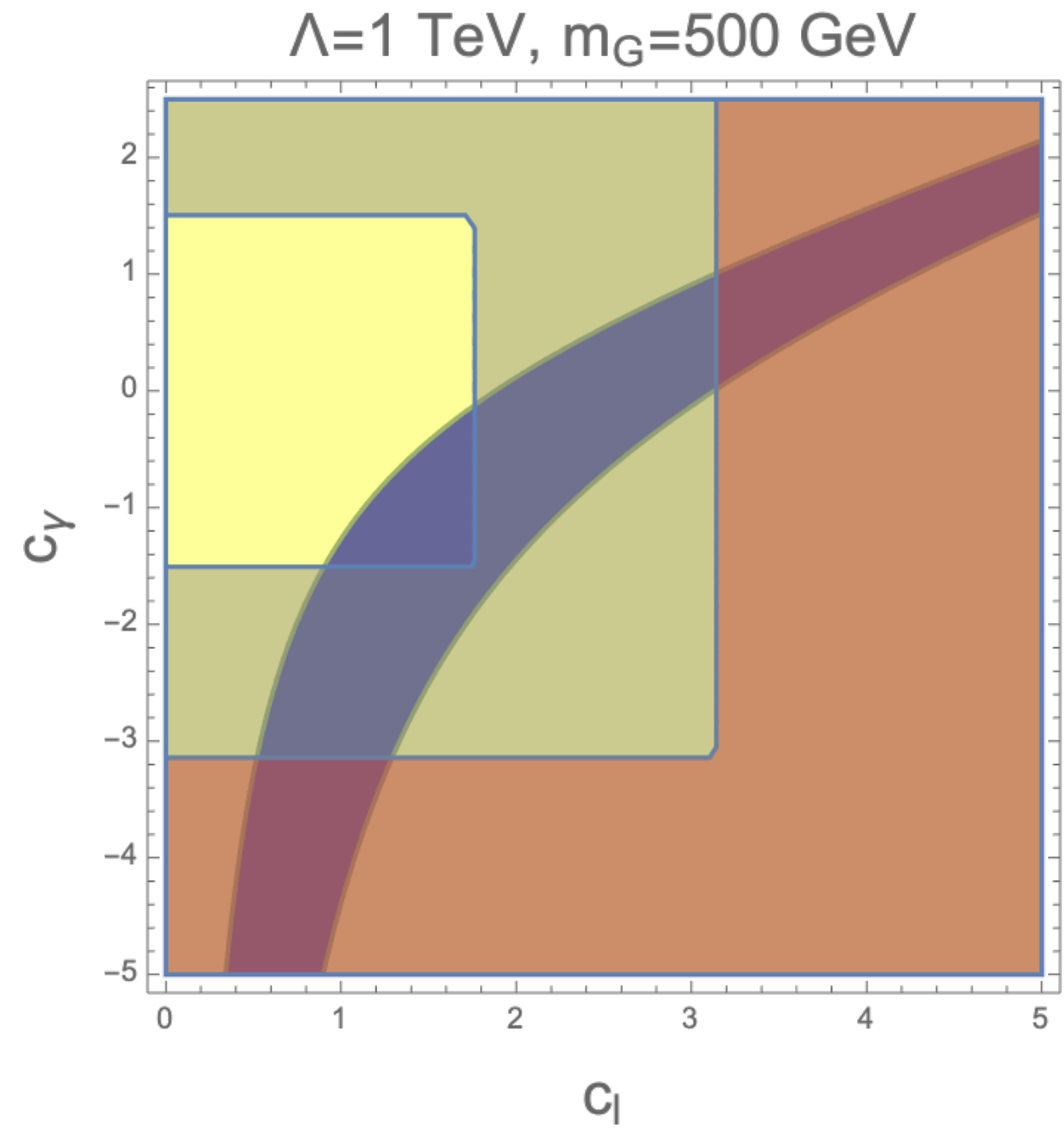}	
	\caption{The parameter space in the $c_\ell$-$c_\gamma$ plane for the cutoff scale fixed at $\Lambda=1$~TeV and the spin-2 particle mass at $m_G = 200$~GeV (left panel) and $500$~GeV (right panel). The blue and yellow shaded regions show the parameter space that can explain the $\Delta a_\mu$ and $\Delta a_e^{\rm LKB}$ anomalies in $2\sigma$ range, while the areas colored in red and gray are excluded by the theoretical constraints from perturbativity and unitarity, respectively.} \label{FigL1000}
\end{figure}

In Fig.~\ref{FigL1000},  the parameter spaces explaining the Muon $(g-2)_\mu$ and LKB $(g-2)_e$ data at $2\sigma$~CL are plotted as the blue and yellow shaded regions in the $c_\ell$-$c_\gamma$ plane, where the cutoff scale is fixed to be $\Lambda =1$~TeV and the massive graviton mass to be $m_G = 200$~GeV (left panel) and 500~GeV (right panel). In the same plots, we also lay out the constraints of the perturbativity and unitarity, and the excluded regions are colored in red and gray, respectively. As a result, it is seen from Fig.~\ref{FigL1000} that, in spite of the strong theoretical perturbativity and unitarity bounds, there is still a substantial portion of parameter space in both plots to explain the $\Delta a_\mu$ and $\Delta a_e^{\rm LKB}$ discrepancies. In particular, it is interesting to note that the muon $g-2$ signal regions are all located in the LBK allowed parameter space, which implies that the current data supports the scenario in which the spin-2 field $G$ couples to all SM charged leptons via a universal coupling, {\it i.e.}, $c_\ell= c_e = c_\mu = c_\tau$. This lepton universality is shown in Ref.~\cite{Huang:2022} to help avoid the strong constraints from the charged-lepton-flavor-violation and $CP$-violation experiments. Furthermore, the comparison of the two plots in Fig.~\ref{FigL1000} indicates that, as the mass of $G$ increases, the unitarity bounds becomes more and more important than the perturbativity ones in limiting the parameter space. Especially, when $m_G = 500$~GeV, the unitarity bounds dominate the theoretical constraint, and shrink the muon $(g-2)$ preferred region to be in a small corner with $1<c_\ell <2$ and $-1.5<c_\gamma<0$.     


\begin{figure}[!htb]
	\centering
	\hspace{-5mm}
	\includegraphics[width=0.48\linewidth]{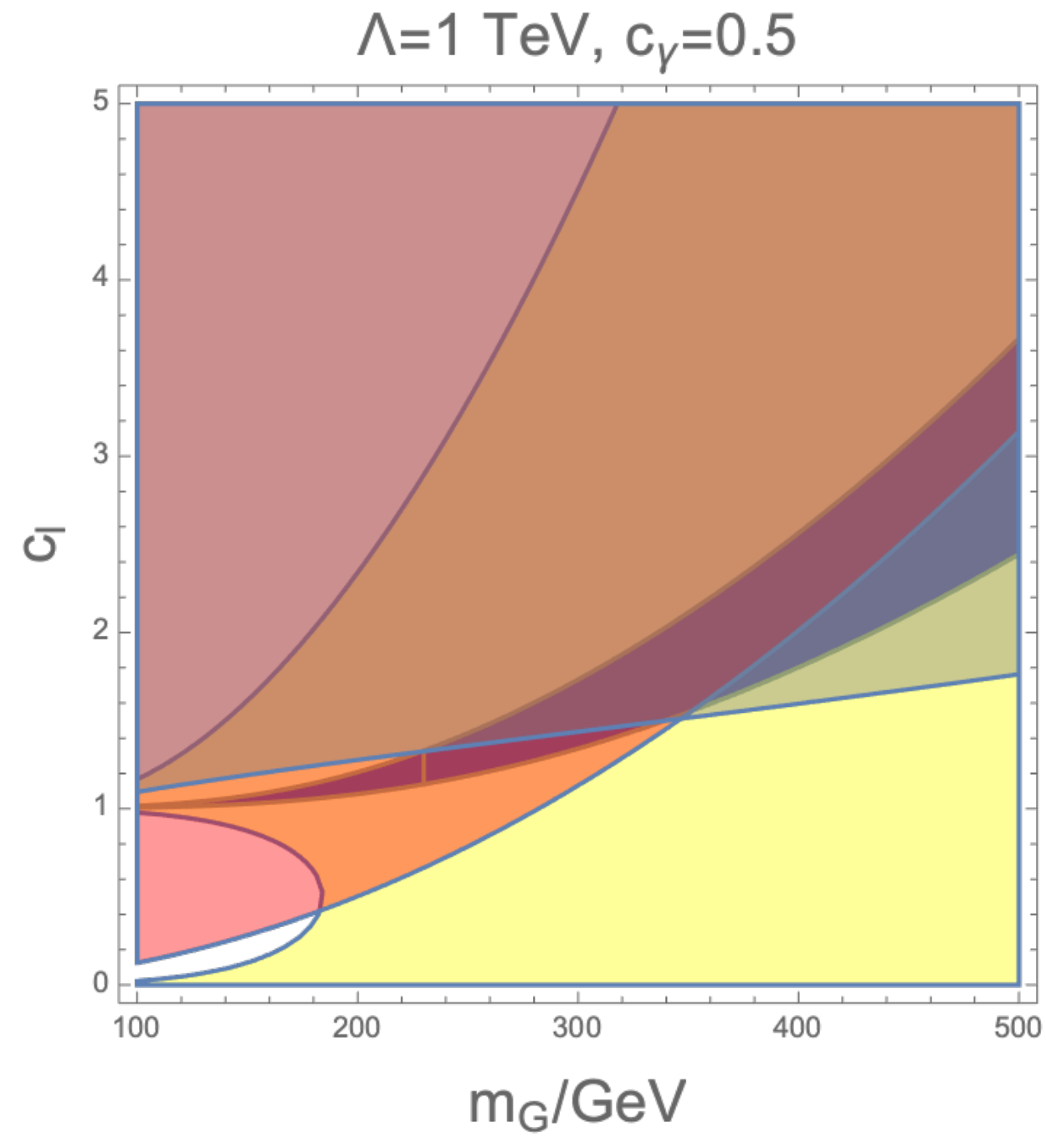}
	\includegraphics[width=0.48\linewidth]{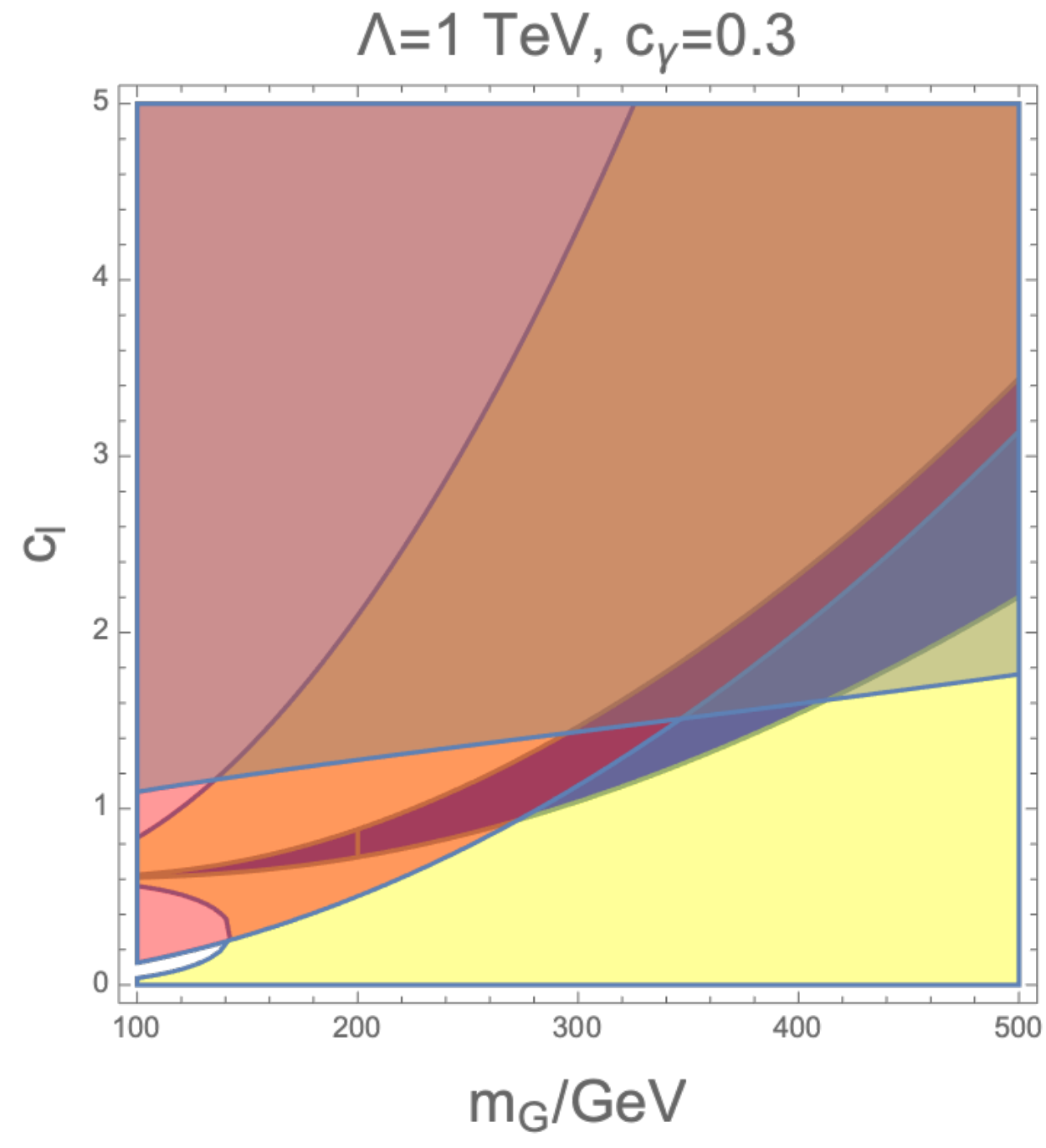}	
	\includegraphics[width=0.48\linewidth]{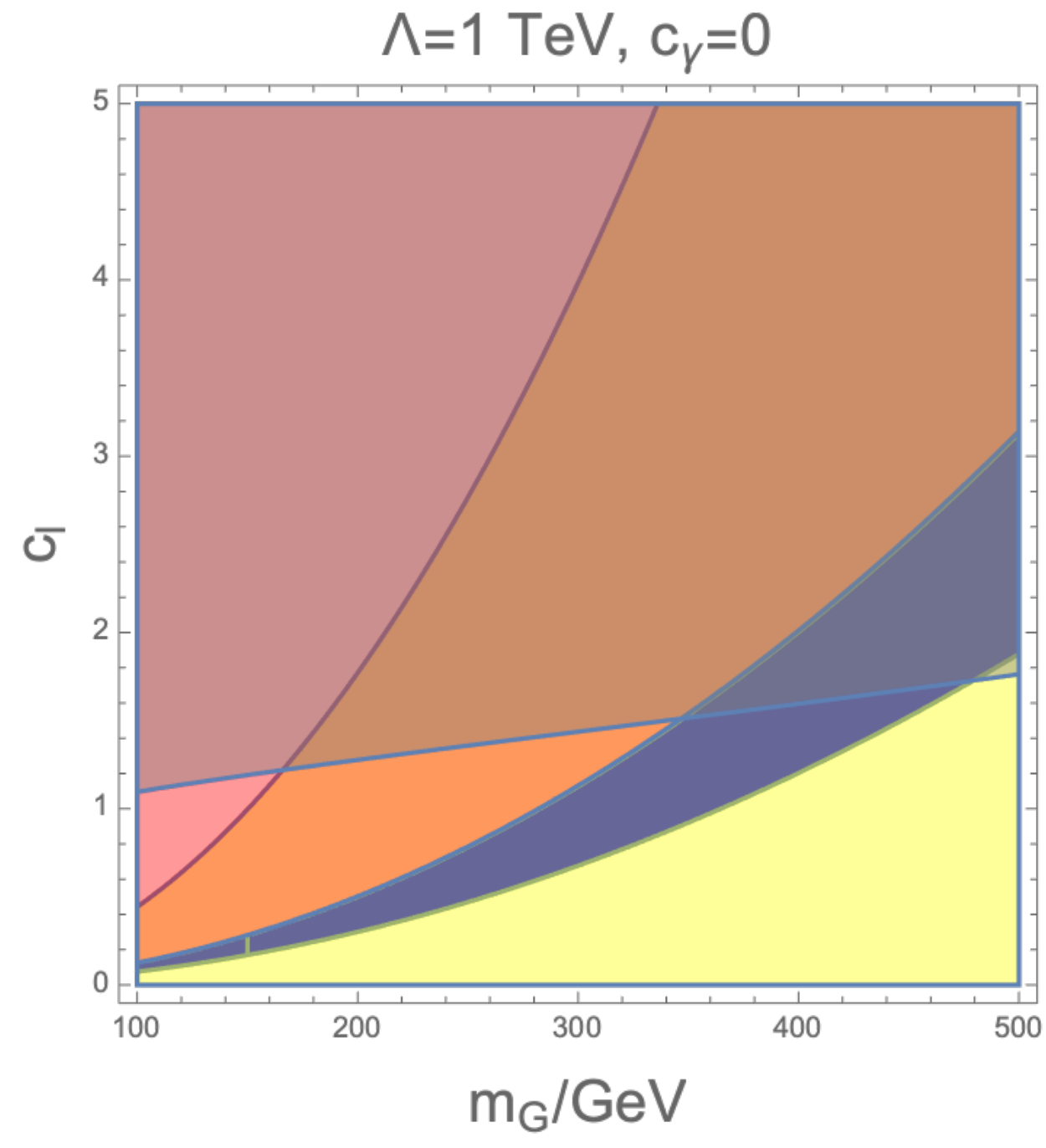}	
	\includegraphics[width=0.48\linewidth]{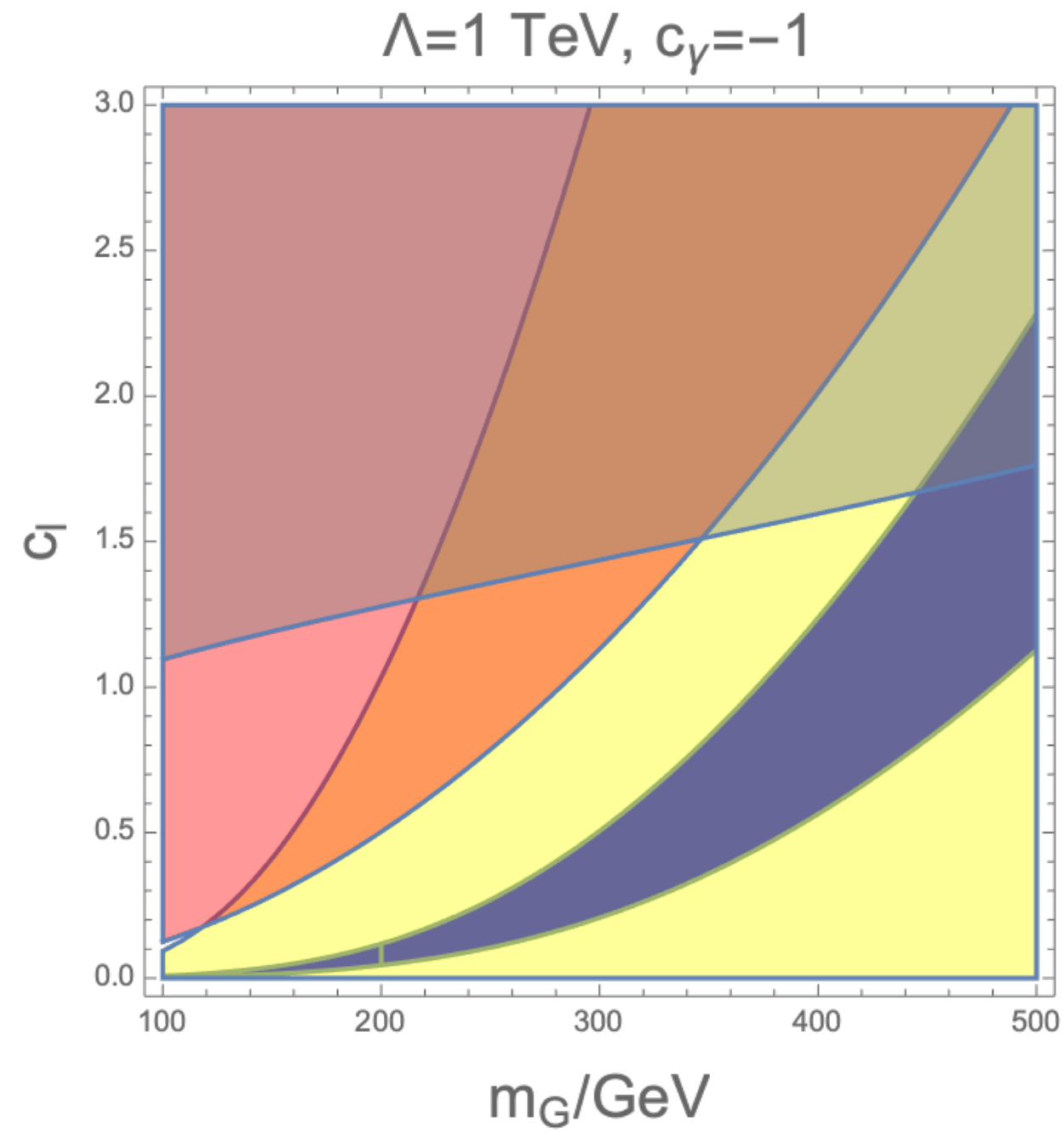}
	\caption{The parameter space in the $m_G$-$c_l$ plane with $\Lambda=1$~TeV and $c_\gamma=0.5$ (top-left panel), 0.3 (top-right panel), 0 (bottom-left panel), and $-1$ (bottom-right panel). The color coding is the same as that in Fig.~\ref{FigL1000} } \label{cGplot}
\end{figure}

We can also show the relevant parameter space in the $m_G$-$c_\ell$ plane as in Fig.~\ref{cGplot}, when the cutoff scale is still fixed to be $\Lambda=1$~TeV and the photon-$G$ coupling is chosen as $c_\gamma = 0.5$, 0.3, 0, and $-1$, respectively. From these plots, it is clear that the unitarity bounds give the stronger constraint than the perturbativity ones in the large spin-2 mass regions. Remarkably, when $c_\gamma\gtrsim 0.5$, the muon $g-2$ signal regions with large $m_G$ values which were open under the perturbativity constraints are now ruled out by the unitarity bounds. However, as the photon-$G$ coupling $c_\gamma$ decreases, more parameter spaces are now allowed by the unitarity. In one special case with $c_\gamma = 0$ where the Barr-Zee-type diagrams~\cite{Bjorken:1977vt,Barr:1990vd} give a vanishing contribution to the lepton $g-2$, the spin-2 particle mass is limited to be below 500~GeV as shown in the lower-left panel in Fig.~\ref{cGplot}. When $c_\gamma$ further drops to take negative values, more allowed parameter regions  now open with $m_G$ extending to even larger values beyond 500~GeV, as shown  in the lower-right panel of Fig.~\ref{cGplot}. 


\begin{figure}[!htb]
	\centering
	\hspace{-5mm}
	\includegraphics[width=0.5\linewidth]{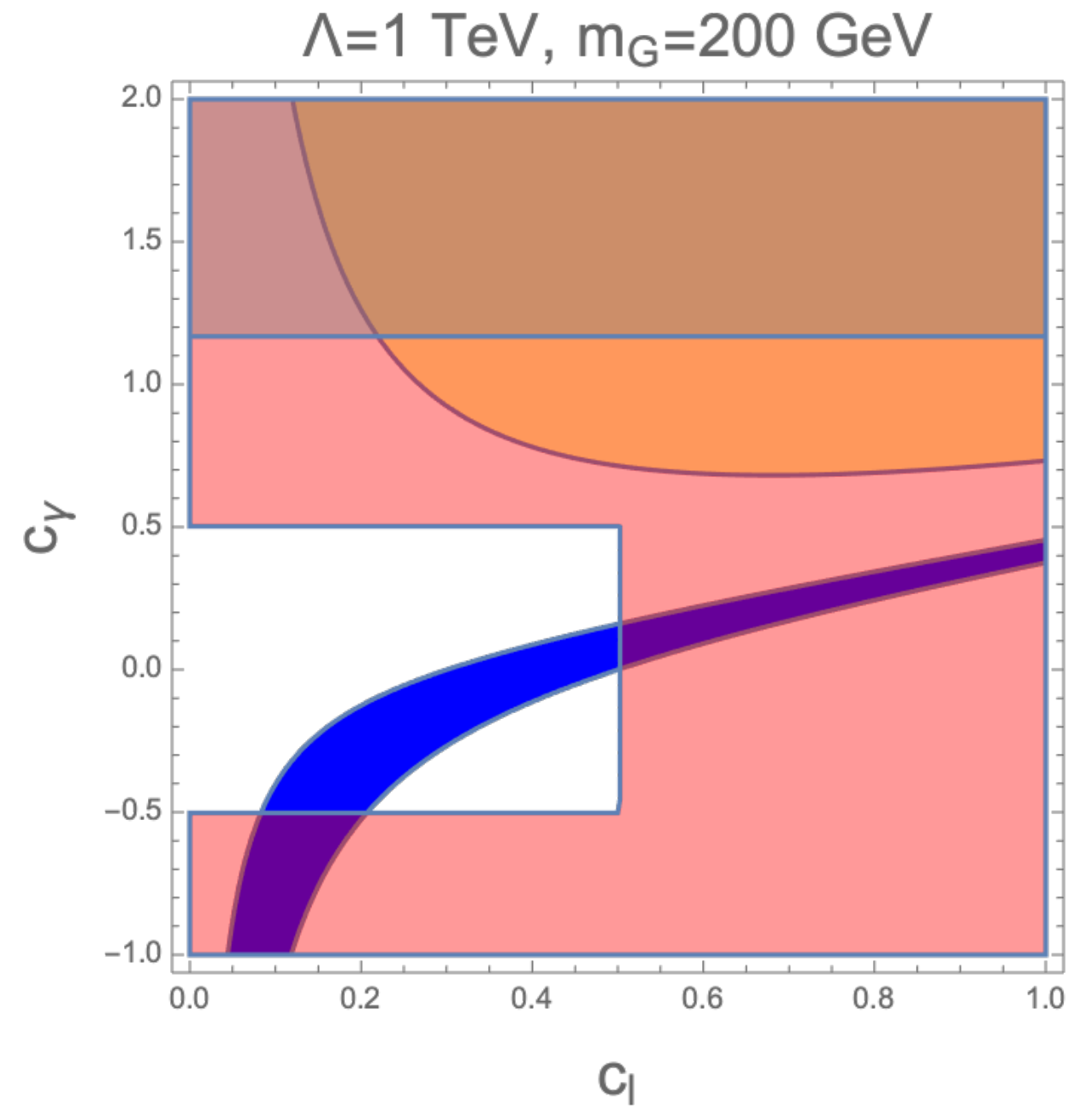}
	\includegraphics[width=0.5\linewidth]{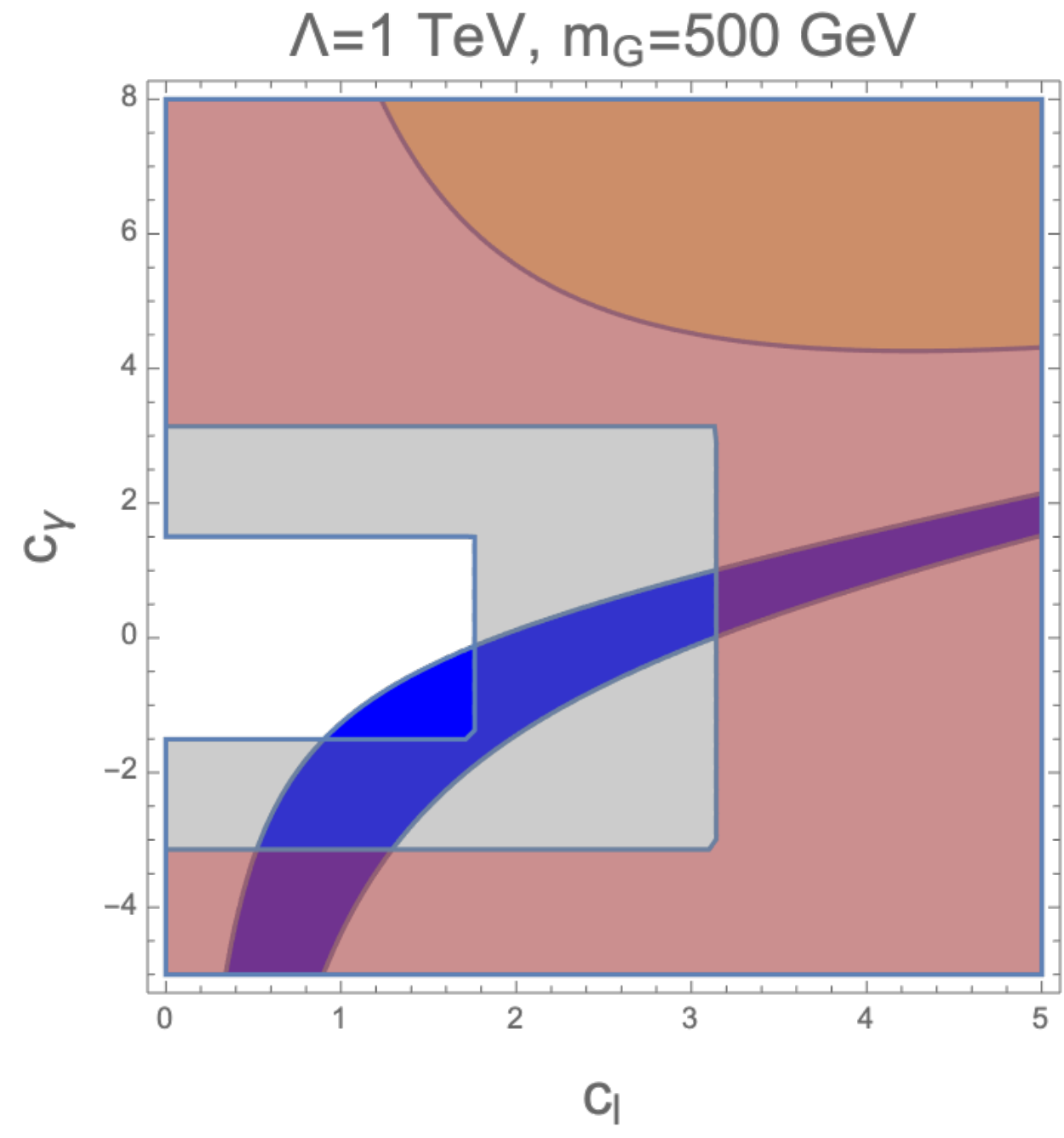}	
	\caption{The parameter space in the $c_\ell$-$c_\gamma$ plane for the cutoff scale fixed at $\Lambda = 1$~TeV and the spin-2 particle mass at $m_G=200$~GeV (left panel) and $500$~GeV (right panel). The color coding is the same as that in Fig.~\ref{FigL1000}, except that the yellow region now represents the parameter space predicted by the Berkeley data of $\Delta a_e^{\rm B}$.} \label{FigBerkeley}
\end{figure}
Finally, we consider the simultaneous interpretation of the $\Delta a_\mu$ anomaly and the Berkeley $\Delta a_e^{\rm B}$ data, 
with the numerical studies in Fig.~\ref{FigBerkeley}. It is remarkable to see that the parameter spaces allowed by the Berkeley $(g-2)_e$ data are completely disfavored by the perturbativity and unitarity constraints. This result can be understood as follows. First of all, it is worth mentioning that the Berkeley data
 prefers a negative $\Delta a_e^{\rm B}$, which is compared with the positive value  $\Delta a_\mu$. Hence, as argued in Ref.~\cite{Huang:2022}, the opposite sign between the muon and electron $g-2$ anomalies can only be achieved by making the Barr-Zee Feynman diagrams~\cite{Bjorken:1977vt,Barr:1990vd} dominate the total lepton $g-2$ formula in Eq.~(\ref{G2total}), since the combination $c_\gamma c_\ell$ in the Barr-Zee contribution can take any sign. On the other hand, the Berkeley result of $\Delta a_e^{\rm B}$ is so large that the coupling $|c_\gamma|$ should take a significant value, which has been strongly ruled out by the perturbativity and unitarity bounds. Especially, in the case with $m_G = 500$~GeV and $\Lambda = 1$~TeV, the unitarity bounds derived in the present paper further strengthen the limits from the perturbativity, which shows the significance of unitarity bounds in constraining our spin-2 particle model.   


\section{Conclusions}\label{Conclusion}
In Ref.~\cite{Huang:2022}, we explain the long-standing discrepancy between the experimental measurements and the SM prediction of the muon $g-2$ in terms of a new massive spin-2 particle, which can be easily derived in the generalized RS models. Note that we have only considered the perturbativity constraints in Ref.~\cite{Huang:2022} on the model parameter space. In the present paper, we have further investigated unitarity bounds on this spin-2 particle model. 
We have obtained the bounds by computing the $s$-wave projected amplitudes for two-body elastic scatterings of charged leptons and photons in the high-energy limit for all possible initial and final helicity states. As a result, we have found that the most stringent constraints on the lepton-$G$ coupling $c_\ell$ and on the photon-$G$ coupling $c_\gamma$ are yielded via the processes $\ell^-_R \ell^+_R \to \ell^-_R \ell^+_R $ and $\gamma_R \gamma_R \to \gamma_R \gamma_R$, respectively. We have applies the obtained unitarity bounds to numerically explore the spin-2 particle parameter space again. It turns out that the tree-level $S$-matrix unitarity gives useful constraints on the muon $g-2$ preferred parameter regions, which is complementary to the perturbativity. 
In particular, it is seen that the unitarity bounds give the strongest limits on $c_\ell$ and $c_\gamma$ in the large $m_G$ region. Nevertheless, despite the stringent constraints imposed by the perturbativity and unitarity, there is still a considerable portion of parameter spaces which can explain the muon $g-2$ anomaly. By further extending this analysis to the electron $g-2$, we have shown that the LKB measurement of $\Delta a_e^{\rm LKB}$ can be compatible with the latest $\Delta a_\mu$ data, while $\Delta a_e^{\rm B}$ obtained at Berkeley cannot be explained by the present spin-2 particle model due to the strong constraints from the perturbativity and unitarity.


\section*{Acknowledgements}
This work is supported in part by the National Key Research and Development Program of China (Grant No. 2020YFC2201501 and No. 2021YFC2203003), the National Natural Science Foundation of China (NSFC)  (Grant No. 12005254 and No. 12147103), and the Key Research Program of Chinese Academy of Sciences  (No. XDPB15).

\appendix



\begin{thebibliography}{0}
\bibitem{pdg}
R.~L.~Workman [Particle Data Group],
PTEP \textbf{2022}, 083C01 (2022)

\bibitem{Muong-2:2006rrc}
G.~W.~Bennett \textit{et al.} [Muon g-2 Collaboration],
Phys. Rev. D \textbf{73}, 072003 (2006)
doi:10.1103/PhysRevD.73.072003
[arXiv:hep-ex/0602035 [hep-ex]].

\bibitem{Muong-2:2021ojo}
B.~Abi \textit{et al.} [Muon g-2 Collaboration],
Phys. Rev. Lett. \textbf{126}, no.14, 141801 (2021)
doi:10.1103/PhysRevLett.126.141801
[arXiv:2104.03281 [hep-ex]].

\bibitem{Aoyama:2020ynm}
T.~Aoyama, \textit{et al.}
Phys. Rept. \textbf{887}, 1-166 (2020)
doi:10.1016/j.physrep.2020.07.006
[arXiv:2006.04822 [hep-ph]].

\bibitem{Keshavarzi:2018mgv}
A.~Keshavarzi, D.~Nomura and T.~Teubner,
Phys. Rev. D \textbf{97}, no.11, 114025 (2018)
doi:10.1103/PhysRevD.97.114025
[arXiv:1802.02995 [hep-ph]].

\bibitem{Chao:2021tvp}
E.~H.~Chao, R.~J.~Hudspith, A.~G\'erardin, J.~R.~Green, H.~B.~Meyer and K.~Ottnad,
Eur. Phys. J. C \textbf{81}, no.7, 651 (2021)
doi:10.1140/epjc/s10052-021-09455-4
[arXiv:2104.02632 [hep-lat]].

\bibitem{Athron:2021iuf}
P.~Athron, C.~Bal\'azs, D.~H.~J.~Jacob, W.~Kotlarski, D.~St\"ockinger and H.~St\"ockinger-Kim,
JHEP \textbf{09}, 080 (2021)
doi:10.1007/JHEP09(2021)080
[arXiv:2104.03691 [hep-ph]].

\bibitem{Huang:2022}
D.~Huang, C.~Q.~Geng and J.~Wu,
[arXiv:2207.13421 [hep-ph]].








\bibitem{Randall:1999ee}
L.~Randall and R.~Sundrum,
Phys. Rev. Lett. \textbf{83}, 3370-3373 (1999)
doi:10.1103/PhysRevLett.83.3370
[arXiv:hep-ph/9905221 [hep-ph]].


\bibitem{Davoudiasl:1999tf}
H.~Davoudiasl, J.~L.~Hewett and T.~G.~Rizzo,
Phys. Lett. B \textbf{473}, 43-49 (2000)
doi:10.1016/S0370-2693(99)01430-6
[arXiv:hep-ph/9911262 [hep-ph]].

\bibitem{Pomarol:1999ad}
A.~Pomarol,
Phys. Lett. B \textbf{486}, 153-157 (2000)
doi:10.1016/S0370-2693(00)00737-1
[arXiv:hep-ph/9911294 [hep-ph]].

\bibitem{Chang:1999nh}
S.~Chang, J.~Hisano, H.~Nakano, N.~Okada and M.~Yamaguchi,
Phys. Rev. D \textbf{62}, 084025 (2000)
doi:10.1103/PhysRevD.62.084025
[arXiv:hep-ph/9912498 [hep-ph]].

\bibitem{Davoudiasl:2000wi}
H.~Davoudiasl, J.~L.~Hewett and T.~G.~Rizzo,
Phys. Rev. D \textbf{63}, 075004 (2001)
doi:10.1103/PhysRevD.63.075004
[arXiv:hep-ph/0006041 [hep-ph]].

\bibitem{Batell:2005wa}
B.~Batell and T.~Gherghetta,
Phys. Rev. D \textbf{73}, 045016 (2006)
doi:10.1103/PhysRevD.73.045016
[arXiv:hep-ph/0512356 [hep-ph]].

\bibitem{Batell:2006dp}
B.~Batell and T.~Gherghetta,
Phys. Rev. D \textbf{75}, 025022 (2007)
doi:10.1103/PhysRevD.75.025022
[arXiv:hep-th/0611305 [hep-th]].

\bibitem{Fok:2012zk}
R.~Fok, C.~Guimaraes, R.~Lewis and V.~Sanz,
JHEP \textbf{12}, 062 (2012)
doi:10.1007/JHEP12(2012)062
[arXiv:1203.2917 [hep-ph]].

\bibitem{Lee:2013bua}
H.~M.~Lee, M.~Park and V.~Sanz,
Eur. Phys. J. C \textbf{74}, 2715 (2014)
doi:10.1140/epjc/s10052-014-2715-8
[arXiv:1306.4107 [hep-ph]].

\bibitem{Han:2015cty}
C.~Han, H.~M.~Lee, M.~Park and V.~Sanz,
Phys. Lett. B \textbf{755}, 371-379 (2016)
doi:10.1016/j.physletb.2016.02.040
[arXiv:1512.06376 [hep-ph]].

\bibitem{Geng:2016xin}
C.~Q.~Geng and D.~Huang,
Phys. Rev. D \textbf{93}, no.11, 115032 (2016)
doi:10.1103/PhysRevD.93.115032
[arXiv:1601.07385 [hep-ph]].

\bibitem{Falkowski:2016glr}
A.~Falkowski and J.~F.~Kamenik,
Phys. Rev. D \textbf{94}, no.1, 015008 (2016)
doi:10.1103/PhysRevD.94.015008
[arXiv:1603.06980 [hep-ph]].

\bibitem{Dillon:2016fgw}
B.~M.~Dillon and V.~Sanz,
Phys. Rev. D \textbf{96}, no.3, 035008 (2017)
doi:10.1103/PhysRevD.96.035008
[arXiv:1603.09550 [hep-ph]].

\bibitem{Dillon:2016tqp}
B.~M.~Dillon, C.~Han, H.~M.~Lee and M.~Park,
Int. J. Mod. Phys. A \textbf{32}, no.33, 1745006 (2017)
doi:10.1142/S0217751X17450063
[arXiv:1606.07171 [hep-ph]].

\bibitem{Kraml:2017atm}
S.~Kraml, U.~Laa, K.~Mawatari and K.~Yamashita,
Eur. Phys. J. C \textbf{77}, no.5, 326 (2017)
doi:10.1140/epjc/s10052-017-4871-0
[arXiv:1701.07008 [hep-ph]].

\bibitem{Geng:2018hpq}
C.~Q.~Geng, D.~Huang and K.~Yamashita,
JHEP \textbf{10}, 046 (2018)
doi:10.1007/JHEP10(2018)046
[arXiv:1807.09643 [hep-ph]].

\bibitem{Goyal:2019vsw}
A.~Goyal, R.~Islam and M.~Kumar,
JHEP \textbf{10}, 050 (2019)
doi:10.1007/JHEP10(2019)050
[arXiv:1905.10583 [hep-ph]].


\bibitem{Gell-Mann:1969cuq}
M.~Gell-Mann, M.~L.~Goldberger, N.~M.~Kroll and F.~E.~Low,
Phys. Rev. \textbf{179}, 1518-1527 (1969)
doi:10.1103/PhysRev.179.1518

\bibitem{Weinberg:1971fb}
S.~Weinberg,
Phys. Rev. Lett. \textbf{27}, 1688-1691 (1971)
doi:10.1103/PhysRevLett.27.1688

\bibitem{Lee:1977yc}
B.~W.~Lee, C.~Quigg and H.~B.~Thacker,
Phys. Rev. Lett. \textbf{38}, 883-885 (1977)
doi:10.1103/PhysRevLett.38.883

\bibitem{Lee:1977eg}
B.~W.~Lee, C.~Quigg and H.~B.~Thacker,
Phys. Rev. D \textbf{16}, 1519 (1977)
doi:10.1103/PhysRevD.16.1519

\bibitem{Durand:1989zs}
L.~Durand, J.~M.~Johnson and J.~L.~Lopez,
Phys. Rev. Lett. \textbf{64}, 1215 (1990)
doi:10.1103/PhysRevLett.64.1215

\bibitem{Glashow:1976nt}
S.~L.~Glashow and S.~Weinberg,
Phys. Rev. D \textbf{15}, 1958 (1977)
doi:10.1103/PhysRevD.15.1958

\bibitem{Huffel:1980sk}
H.~Huffel and G.~Pocsik,
Z. Phys. C \textbf{8}, 13 (1981)
doi:10.1007/BF01429824

\bibitem{Maalampi:1991fb}
J.~Maalampi, J.~Sirkka and I.~Vilja,
Phys. Lett. B \textbf{265}, 371-376 (1991)
doi:10.1016/0370-2693(91)90068-2

\bibitem{Kanemura:1993hm}
S.~Kanemura, T.~Kubota and E.~Takasugi,
Phys. Lett. B \textbf{313}, 155-160 (1993)
doi:10.1016/0370-2693(93)91205-2
[arXiv:hep-ph/9303263 [hep-ph]].

\bibitem{Akeroyd:2000wc}
A.~G.~Akeroyd, A.~Arhrib and E.~M.~Naimi,
Phys. Lett. B \textbf{490}, 119-124 (2000)
doi:10.1016/S0370-2693(00)00962-X
[arXiv:hep-ph/0006035 [hep-ph]].

\bibitem{Das:2015mwa}
D.~Das and I.~Saha,
Phys. Rev. D \textbf{91}, no.9, 095024 (2015)
doi:10.1103/PhysRevD.91.095024
[arXiv:1503.02135 [hep-ph]].

\bibitem{Kanemura:2015ska}
S.~Kanemura and K.~Yagyu,
Phys. Lett. B \textbf{751}, 289-296 (2015)
doi:10.1016/j.physletb.2015.10.047
[arXiv:1509.06060 [hep-ph]].

\bibitem{Goodsell:2018tti}
M.~D.~Goodsell and F.~Staub,
Eur. Phys. J. C \textbf{78}, no.8, 649 (2018)
doi:10.1140/epjc/s10052-018-6127-z
[arXiv:1805.07306 [hep-ph]].

\bibitem{Appelquist:1987cf}
T.~Appelquist and M.~S.~Chanowitz,
Phys. Rev. Lett. \textbf{59}, 2405 (1987)
[erratum: Phys. Rev. Lett. \textbf{60}, 1589 (1988)]
doi:10.1103/PhysRevLett.59.2405


\bibitem{Chaichian:1987zt}
M.~Chaichian and J.~Fischer,
Nucl. Phys. B \textbf{303}, 557-568 (1988)
doi:10.1016/0550-3213(88)90394-X


\bibitem{Banta:2021dek}
I.~Banta, T.~Cohen, N.~Craig, X.~Lu and D.~Sutherland,
JHEP \textbf{02}, 029 (2022)
doi:10.1007/JHEP02(2022)029
[arXiv:2110.02967 [hep-ph]].

\bibitem{Han:1998sg}
T.~Han, J.~D.~Lykken and R.~J.~Zhang,
Phys. Rev. D \textbf{59}, 105006 (1999)
doi:10.1103/PhysRevD.59.105006
[arXiv:hep-ph/9811350 [hep-ph]].

\bibitem{Wu:2002xa}
Y.~L.~Wu,
Int. J. Mod. Phys. A \textbf{18}, 5363-5420 (2003)
doi:10.1142/S0217751X03015222
[arXiv:hep-th/0209021 [hep-th]].

\bibitem{Wu:2003dd}
Y.~L.~Wu,
Mod. Phys. Lett. A \textbf{19}, 2191-2204 (2004)
doi:10.1142/S0217732304015361
[arXiv:hep-th/0311082 [hep-th]].

\bibitem{Nebot:2007bc}
M.~Nebot, J.~F.~Oliver, D.~Palao and A.~Santamaria,
Phys. Rev. D \textbf{77}, 093013 (2008)
doi:10.1103/PhysRevD.77.093013
[arXiv:0711.0483 [hep-ph]].

\bibitem{Morel:2020dww}
L.~Morel, Z.~Yao, P.~Clad\'e and S.~Guellati-Kh\'elifa,
Nature \textbf{588}, no.7836, 61-65 (2020)
doi:10.1038/s41586-020-2964-7

\bibitem{Parker:2018vye}
R.~H.~Parker, C.~Yu, W.~Zhong, B.~Estey and H.~M\"uller,
Science \textbf{360}, 191 (2018)
doi:10.1126/science.aap7706
[arXiv:1812.04130 [physics.atom-ph]].

\bibitem{Aoyama:2012wj}
T.~Aoyama, M.~Hayakawa, T.~Kinoshita and M.~Nio,
Phys. Rev. Lett. \textbf{109}, 111807 (2012)
doi:10.1103/PhysRevLett.109.111807
[arXiv:1205.5368 [hep-ph]].

\bibitem{Aoyama:2019ryr}
T.~Aoyama, T.~Kinoshita and M.~Nio,
Atoms \textbf{7}, no.1, 28 (2019)
doi:10.3390/atoms7010028

\bibitem{Hanneke:2008tm}
D.~Hanneke, S.~Fogwell and G.~Gabrielse,
Phys. Rev. Lett. \textbf{100}, 120801 (2008)
doi:10.1103/PhysRevLett.100.120801
[arXiv:0801.1134 [physics.atom-ph]].

\bibitem{Bjorken:1977vt}
J.~D.~Bjorken and S.~Weinberg,
Phys. Rev. Lett. \textbf{38}, 622 (1977)
doi:10.1103/PhysRevLett.38.622

\bibitem{Barr:1990vd}
S.~M.~Barr and A.~Zee,
Phys. Rev. Lett. \textbf{65}, 21-24 (1990)
[erratum: Phys. Rev. Lett. \textbf{65}, 2920 (1990)]
doi:10.1103/PhysRevLett.65.21




\end{thebibliography}
\end{document}